\documentclass[aip,amsmath,amssymb,reprint,floatfix]{revtex4-1}

\usepackage[utf8]{inputenc}
\usepackage{xcolor}
\usepackage[colorlinks, linkcolor=blue, citecolor=blue, urlcolor=blue]{hyperref}
\usepackage{graphicx}
\usepackage{physics}
\usepackage{siunitx}
\usepackage{bbold}
\usepackage{booktabs}
\usepackage{multirow}
\usepackage{longtable}
\usepackage{lipsum}

\newcommand{\mi}{ {\rm i} }

\newcommand{\id}{\mathbb{1}}

\begin{document}

\title{A quantum optics approach to photoinduced electron transfer in cavities}

\author{D.~Wellnitz}
\thanks{david.wellnitz@etu.unistra.fr}
\affiliation{ISIS (UMR 7006), University of Strasbourg and CNRS, and icFRC, 67000 Strasbourg, France}
\affiliation{IPCMS (UMR 7504), CNRS, 67000 Strasbourg, France}

\author{G.~Pupillo}
\thanks{pupillo@unistra.fr}
\affiliation{ISIS (UMR 7006), University of Strasbourg and CNRS, and icFRC, 67000 Strasbourg, France}

\author{J.~Schachenmayer}
\thanks{schachenmayer@unistra.fr}
\affiliation{ISIS (UMR 7006), University of Strasbourg and CNRS, and icFRC, 67000 Strasbourg, France}
\affiliation{IPCMS (UMR 7504), CNRS, 67000 Strasbourg, France}
\date{February 2021}

\begin{abstract}
    We study a simple model for photoinduced electron transfer reactions for the case of many donor-acceptor pairs that are collectively and homogeneously coupled to a photon mode of a cavity. We describe both coherent and dissipative collective effects resulting from this coupling within the framework of a quantum optics Lindblad master equation. We introduce a method to derive an effective rate equation for electron transfer, by adiabatically eliminating donor and acceptor states and the cavity mode. The resulting rate equation is valid both for weak and strong coupling to the cavity mode, and describes electronic transfer through both the cavity coupled bright states and the uncoupled dark states. We derive an analytic expression for the instantaneous electron transfer rate that depends non-trivially on the time-varying number of pairs in the ground state. We find that under proper resonance conditions, and in the presence of an incoherent drive, reaction rates can be enhanced by the cavity. This enhancement persists, and can even be largest, in the weak light-matter coupling regime. We discuss how the cavity effect is relevant for realistic experiments.
\end{abstract}

\maketitle

\section{Introduction}

Collectively coupling a large number of molecules to a cavity mode can lead to a significant modification of chemical reaction rates~\cite{Ebbesen_Hybrid_2016,Herrera2020Mar}. This has been demonstrated in a series of recent breakthrough experiments where vibrational~\cite{Thomas2016Sep,Dunkelberger2016Nov,Thomas_Tiltin_2019,Lather2019Jul,Hiura2019Sep} or electronic~\cite{Hutchison_Modify_2012,Munkhbat2018Jul,Peters2019Mar} transitions have been strongly coupled to cavity modes or other confined electromagnetic fields. Theoretically understanding such collective cavity-modified chemistry is a major challenge due to the difficulty of modeling the dynamics of large  ensembles of molecules with many internal (i.e., electronic, vibrational, etc.) and motional degrees of freedom together with those of the electromagnetic field. Cavity-induced effects in polaritonic chemistry have often been described within a coherent Hamiltonian framework~\cite{herrera2016cavity,kowalewski2016cavity,Flick2017Mar,feist2018polaritonic,Flick2018Sep,vendrell2018collective,campos2019resonant}, however incoherent dissipation from molecular radiative and non-radiative transitions, different molecular environments, cavity losses, etc., can play an important role in the dynamics~\cite{Fregoni2020Jan,Herrera2020Jan,Ulusoy2020Jul,Felicetti2020Sep,Wang2020Feb,Antoniou2020Sep,Davidsson2020Oct,Torres-Sanchez2020Nov}. In particular, cavity losses represent an ultrafast (fs timescale) decay mechanism unique to polaritonic chemistry~\cite{Herrera2017Nov}. The interplay between coherent Hamiltonian dynamics and these dissipation channels adds an additional layer of complexity for theory.

In order to understand basic mechanisms underlying cavity-modified chemistry, it can be thus instructive to analyze simplified models of few-level systems, which can be efficiently described by standard quantum optics tools~\cite{delPino2015May,Martinez-Martinez2018Mar,Reitz2019May,Fregoni2020Jan,Herrera2020Jan}. For example, in many situations where a large dissipation is present one can use an adiabatic elimination procedure~\cite{Finkelstein-Shapiro2020Adiabatic,Reiter2012Mar}, a theoretical tool that can make it possible to significantly reduce the complexity of the problem by finding analytical solutions to the molecular dynamics of just a few key degrees of freedom~\cite{Wellnitz2020Feb}. Using these techniques for the case of ground-state atoms cooled to submillikelvin temperatures~\cite{Jin2012Sep,Dulieu2017Dec}, we have recently shown that collective dissipative effects can compete with, and even dominate over, coherent Hamiltonian dynamics for molecular formation in a cavity in realistic experiments with up to $10^5$ molecules -- a dissipative form of polaritonic chemistry~\cite{Wellnitz2020Feb}. It is an interesting question to explore to what extent these dissipative mechanisms and theoretical techniques can be used to investigate reactions in more ordinary situations in chemistry.

\smallskip

Here, we study the Lindblad master equation dynamics that describes a photoinduced electron transfer reaction of many donor-acceptor pairs homogeneously coupled to a cavity. We simplify each pair to a 4-level system, an approximation which neglects internal and external motional degrees of freedom~\cite{Mandal_Polari_2020}. In our model an electron donor is excited by an incoherent external field, leaving a single electron loosely bound. This electron is then coherently transferred to an acceptor, which finally relaxes incoherently into a final state. We introduce and explain the concept of the adiabatic elimination procedure as a tool to analyze dissipative polariton chemistry, and clarify under which conditions it may be used. In order to obtain an analytical result for the modification of the transfer rate from the donor to the acceptor, we adiabatically eliminate the cavity mode and the excited donor/acceptor states. We show that for our specific problem this leads to a purely dissipative, and essentially classical, effective master equation without any coherent contribution, that is, with effective Hamiltonian
\begin{align}
    \hat H_{\rm eff} = 0 \,.
\end{align}
The master equation can be numerically simulated for a very large number of donor-acceptor pairs, and makes it possible to derive an analytical expression for the instantaneous transfer rate, given in Eq.~\eqref{eq:transrate}. The coherent (quantum) dynamics of the excited states is implicitly included in this transfer rate, capturing the effects of virtually excited polaritons (in the strong coupling regime), superradiant states (in the weak coupling regime), and dark states, all within a single formula. We analyze the validity of the analytical formula and discuss the conditions under which it can describe observations in realistic experimental setups.

\smallskip

We find that electron transfer occurs via two distinct types of channels: One type comprises $N$ transfer channels that are activated by spontaneous absorption of an incoherent photon by an individual donor-acceptor pair. For large $N$, these transfer channels use mostly ``dark states'' that are decoupled from the cavity, and they are therefore essentially independent of the cavity coupling strength $g$. The other transfer channel is activated by absorption of an incoherent photon into the cavity. Here, the electron transfer occurs via the collective, cavity coupled states, in particular super-radiant or polaritonic states, which we call ``bright states''. This channel is added by the cavity and has a strong dependence on $g$. The transfer efficiency is maximal for an intermediate cavity coupling that can be in the weak or strong coupling regime, depending on the model parameters. We find that the transfer rate scales non-trivially with increasing number of donor-acceptor pairs and that for realistic situations the cavity transfer channel can dominate and enhance the transfer rate.

\smallskip

Those results are in line with recent theoretical studies for small system sizes~\cite{Fregoni2020Jan,Herrera2020Jan,Ulusoy2020Jul,Felicetti2020Sep,Wang2020Feb,Antoniou2020Sep,Davidsson2020Oct,Torres-Sanchez2020Nov} that have highlighted the role of dissipation, and possibilities for optimal coupling strengths close to the weak coupling regime. In contrast to previous works, here we confirm such observations in toy-models for macroscopic molecule numbers. Further, recent research has highlighted that collective effects lead to a modified reaction dynamics that differs from a simple increase of the Rabi splitting with $N$, in a coherent Hamiltonian framework without dissipation~\cite{Ulusoy2019Oct,Szidarovszky2020May,Davidsson2020Jun}. Here we derive a non-linear rate equation for a purely dissipative regime, and find that instantaneous transfer rates are non-trivial due to a combination of transfer through dark and polaritonics states, which depends on the instantaneous number of coupled pairs, $N(t)$. We note that while in our work the role of dark and bright states depends directly on the relative rates at which they are externally excited, the natural mixing of dark and bright states due to disorder and vibrational couplings has been also investigate in several recent works~\cite{Gonzalez-Ballestero2016Oct,vendrell2018collective,Groenhof2019Sep,Sommer2020Oct,Botzung2020Oct,Chavez2020Oct}.

\smallskip

The remainder of the paper is organized as follows: We start by introducing our simplified electron transfer model in Sec.~\ref{sec:model}. There, we explain how to describe both coherent and dissipative processes within the master equation approach in Sec.~\ref{ssec:master}, and discuss the relevance of this model to realistic experiments in Sec.~\ref{ssec:nanocrystalnumbers}. We then present the adiabatic elimination procedure in detail in Sec.~\ref{sec:elim}. In Sec.~\ref{sec:results} we discuss the effective master equation: In Sec.~\ref{ssec:rateeq} we provide an analytical expression for the electron transfer rate and clarify the contributions of the different transfer channels; in Sec.~\ref{ssec:evol} we demonstrate the validity of our rate equation and analyze the resulting dynamics; and in Sec.~\ref{ssec:discussion} we evaluate the parameter dependencies of the transfer rate, and estimate when the cavity can increase it. We provide a conclusion and an outlook in Sec.~\ref{sec:conclusion}.

\section{Model}
\label{sec:model}

\begin{figure}
    \centering
    \includegraphics[width=\columnwidth]{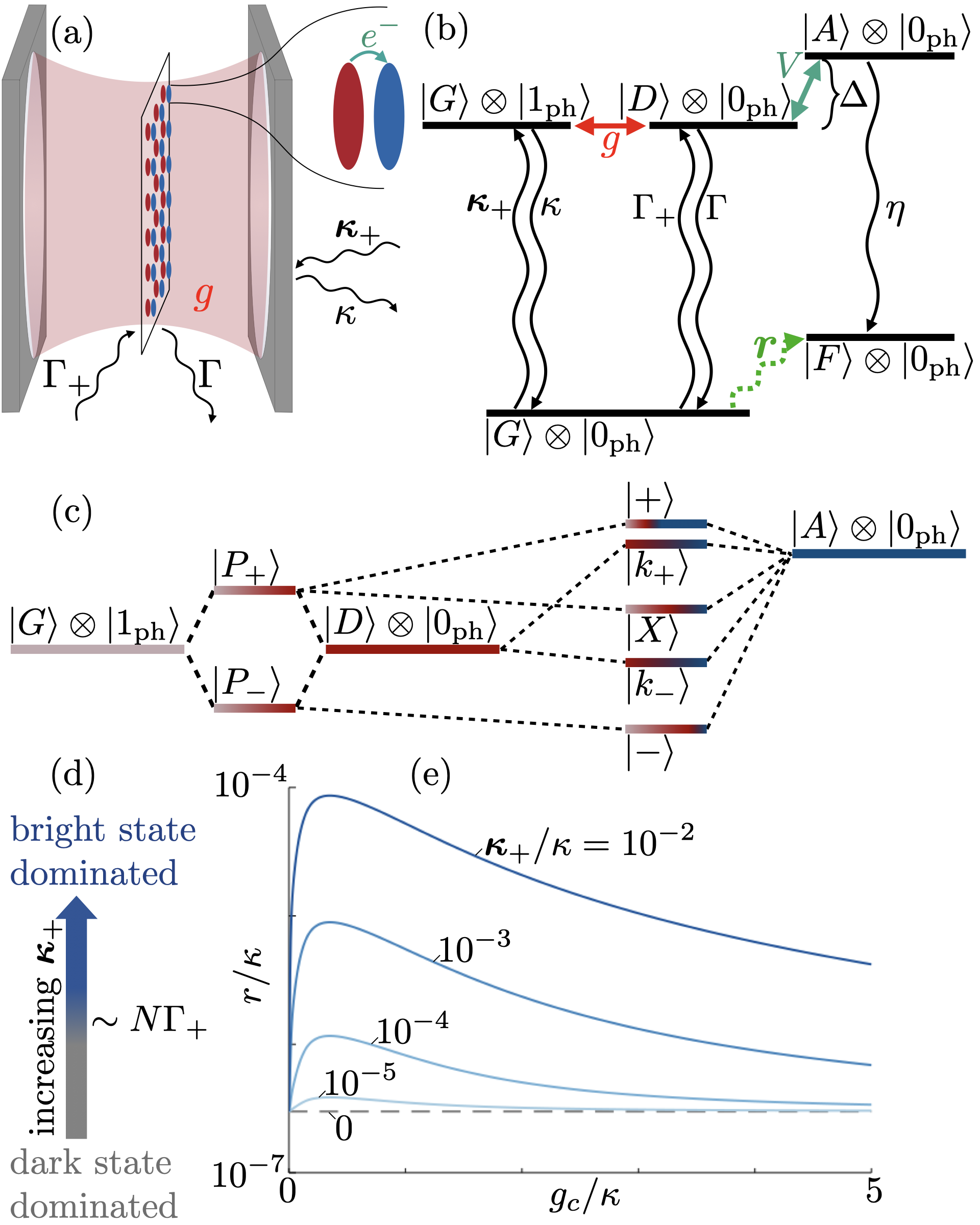}
    \caption{
    (a) Schematic model. $N$ donor (red) - acceptor (blue) pairs are homogeneously coupled to a cavity with coupling constant $g$. Each pair and the cavity are incoherently pumped at rate $\Gamma_+$ and $\kappa_+$, and decay at rate $\Gamma$ and $\kappa$, respectively. 
    (b) Level scheme of a single molecule coupled to the cavity: Each pair is described as a 4-level system with a ground state $\ket G$, excited donor ($\ket D$) and acceptor state ($\ket A$), and a final state $\ket F$, corresponding to the reaction product. The zero and one photon-number states of the cavity are denoted as $\ket{0_{\rm ph}}$ and $\ket{1_{\rm ph}}$, respectively. 
    The cavity coherently couples the states $\ket{G} \otimes \ket{1_{\rm ph}}$ and $\ket{D}\otimes\ket{0_{\rm ph}}$ with strength $g$, while coherent coupling between $\ket{D}\otimes\ket{0_{\rm ph}}$ and $\ket{A}\otimes \ket{0_{\rm ph}}$ with strength $V$ and detuning $\Delta$ induces the electron transfer. After the adiabatic elimination, all coherent and dissipative dynamics can be captured by an effective transfer rate $r$ of population from $\ket G$ to $\ket F$ (green dotted arrow).
    (c) Schematic level-diagram for eigenstates of Hamiltonian Eq.~\eqref{eq:ham} (not to scale).
    (d) For increasing $\kappa_+$, the dynamics transitions from being dominated by dark states to being dominated by bright states above $\kappa_+ \gtrsim N\Gamma_+$.
    (e) Transfer rate $r$ from Eq.~\eqref{eq:transrate} on a logarithmic scale as a function of the collective cavity coupling $g_c = \sqrt{N}g$ for various cavity pumping rates $0 \leq \kappa_+ \leq 10^{-2}\kappa$ with $N=10^4$, $\Gamma=3\times10^{-7}\kappa$, $\Gamma_+ = 10^{-4}\Gamma$, $\Delta=0.2\kappa$, $V=0.1\kappa$, and $\eta = 10^{-2}\kappa$.}
    \label{fig:setup}
\end{figure}

\subsection{Quantum optics master equation approach}
\label{ssec:master}

We consider a model of $N$ donor-acceptor pairs homogeneously coupled to a single-mode cavity as schematically depicted in Fig.~\ref{fig:setup}(a). Each donor-acceptor pair is described by a 4-level system with a ground state $\ket G$, an excited donor state $\ket D$, an excited acceptor state $\ket A$, and a final state $\ket F$; the cavity is described by a photon mode with annihilation operator $\hat a$.  We are interested in the transfer dynamics from $\ket G$ to $\ket F$. The cavity is resonantly coupled to the $\ket G \leftrightarrow \ket D$ transition with single molecule coupling strength $g$. We further include the following dissipative processes: incoherent cavity pumping at rate $\kappa_+$, leading to the creation of a cavity photon, cavity decay at rate $\kappa$, leading to loss of photons from the cavity, pumping of the individual pairs from their ground state $\ket G$ to donor state $\ket D$ at rate $\Gamma_+$, the inverse decay process $\Gamma$, and an artificially introduced non-radiative relaxation from the acceptor state $\ket A$ to the final state $\ket F$ at rate $\eta$. The corresponding level scheme including the cavity states $\ket{0_{\rm ph}}$ and $\ket{1_{\rm ph}}$ with zero and one photon, respectively, is shown in Fig~\ref{fig:setup}(b).

We set $\hbar = 1$ and work in a rotating frame
\footnote{Precisely, the rotating frame transformation is achieved by applying the unitary operator $\hat U = \exp[\mi \omega_{DG}t \hat N_e]$, where the Hamiltonian and the states transform according to $\hat H' = \hat U \hat H \hat U^\dagger + \mi (\partial_t\hat U) \hat U^\dagger$ and $\ket{\psi'} = \hat U \ket \psi$, respectively.}
where we include the bare excited state energies derived from the operator $\hat N_e\omega_{DG} + \hat N_F \omega_{FG} = \hat N_e (E_D - E_G) + \hat N_F (E_F - E_G)$ in the definition of the states. Here, the operator $\hat N_e = \hat N_D + \hat N_A + \hat a^\dagger \hat a$ gives the excitation number, with the population number operators $\hat N_\psi = \sum_n \ket \psi \bra \psi_n$, $\psi \in \{G,D,A,F\}$, and the subscript $n$ labeling the $n$th pair. We also make the rotating wave approximation, and neglect all counter-rotating terms, which is valid for $g_c \equiv \sqrt{N}g \ll \omega_{DG}$~\cite{Mandal_Polari_2020}. Under these considerations, all coherent dynamics is described by the Hamiltonian
\begin{align}
    \hat H = &\Delta \hat N_A + \sum_{n=1}^N V \qty(\ket D \bra A_n + \ket A \bra D_n) \nonumber \\
    &+ \sum_{n=1}^N g \qty(\ket D \bra G_n \hat a + \ket G \bra D_n \hat a^\dagger) \label{eq:ham} \, .
\end{align}
The first line on the right hand side of Eq.~\eqref{eq:ham} describes the contributions due to the energy difference $\Delta = E_A - E_D$ between the states $\ket A$ and $\ket D$, and the coherent coupling of the two states with strength $V$ associated with the tunneling of an electron from the donor to the acceptor. The second line describes Tavis-Cummings like coupling between the transition $\ket G \leftrightarrow \ket D$ and the cavity with strength $g$, in particular the absorption of a cavity photon by a pair in state $\ket G$, bringing it to the excited donor state $\ket D$, and the inverse process. This Hamiltonian part gives rise to two upper and lower polariton states $\ket{P_\pm}$ at energies $\pm g_c$. This term induces an indirect symmetric coupling between the pairs via the cavity, leading to collective effects.

The Hamiltonian in Eq.~\eqref{eq:ham} commutes with $\hat N_e$ and thus cannot change the number of excitations. In the following we discuss the eigenstates of $\hat H$ in the single excitation limit $\langle \hat N_e \rangle \leq 1$ [see Fig.~\ref{fig:setup}(c) for a level-diagram sketch]. Ignoring $\ket F$ for the moment, the total ground state $\ket{G_c}$ of the system ($\langle \hat N_e \rangle = 0$) is the state with all individual pairs in state $\ket G$, and no cavity photons. In the first excited manifold ($\langle\hat N_e\rangle = 1$), we find $2N-2$ dark states, labelled as $\ket{k_\pm}$, with zero photon weight and uncoupled from the cavity, as ensured by the condition $(\sum_n \ket G \bra D_n) \ket{k_\pm} = \hat a \ket{k_\pm} = 0$. The coherent coupling between the states $\ket D$ and $\ket A$ induces a separation of the dark states into two sets of $N-1$ energetically degenerate states at energies $\Delta/2 \pm \sqrt{\Delta^2/4 + V^2}$. The label $k\neq 0$ indicates their $N-1$ ``quasi-momenta'' defined by a discrete Fourier transform, which means that there is a phase $\exp(-2\pi \mi kn/N)$ for the $n$th pair in an excited state (see Appendix~\ref{app:eigenstates} for their precise definition). The three remaining eigenstates of Eq.~\eqref{eq:ham}  are those states that couple to light. These states are labeled $\ket +$, $\ket -$, and $\ket X$, they are superpositions of $\ket{P_\pm}$ and $\sum_n \ket{A}_n/\sqrt{N}$, where $\ket{A}_n \equiv (\ket{A}\bra{G}_n) \ket{G_c}$ labels the state with pair $n$ in state $\ket A$, and can be obtained by diagonalizing the remaining $3\times 3$ block matrix in the Hamiltonian. Concerning the donor-acceptor pairs in state $\ket F$, we note that those pairs do not add any coherent contribution in the Hamiltonian Eq.~\eqref{eq:ham}. We can thus ignore them and construct the eigenstates as given above for a reduced number of pairs (see Appendix~\ref{app:eigenstates}). Note that in our model we neglect disorder or vibrational couplings which would lead to a mixing of polariton states $\ket{\pm,X}$ with dark states $\ket{k_\pm}$ and give rise to effective coherent mixing and decay processes between those states~\cite{Gonzalez-Ballestero2016Oct,vendrell2018collective,Groenhof2019Sep,Sommer2020Oct,Botzung2020Oct,Chavez2020Oct}.

In order to model both coherent and dissipative dynamics, we use a Lindblad master equation~\cite{gardiner2004quantum}
\begin{align}
    \partial_t \hat \rho = -\mi \qty[\hat H, \hat \rho] + \sum_k \mathcal D \qty[\hat L_k] \hat \rho \label{eq:master} \, ,
\end{align}
with the dissipator $\mathcal D [\hat L] \hat \rho \equiv - \hat L^\dagger \hat L \hat \rho - \hat \rho \hat L^\dagger \hat L + 2 \hat L \hat \rho \hat L^\dagger$, see below. The density matrix $\hat \rho$ is a hermitian matrix capturing all state populations on the diagonal and coherences between states on the off-diagonal. The sum runs over all decay channels, describing the dissipative processes characterized by Lindblad operators $\hat L_k$, induced by the exchange of energy between the system and a much larger ``bath''. Equation~\eqref{eq:master} is a valid description if the bath that the energy is transferred to is ``Markovian'' --- that is, it thermalizes at a timescale much faster than the timescale associated with any coherent coupling back to the system. This is typically true for the background electromagnetic field in our setup. The incoherent decay processes considered here are described by $2+3N$ Lindblad operators
\begin{align}
    \hat L_{\kappa_+} &= \sqrt{\frac{\kappa_+}{2}} \hat a^\dagger \,,\\
    \hat L_\kappa &= \sqrt{\frac{\kappa}{2}} \hat a \,,\\
    \hat L_{\Gamma_+}^{(n)} &= \sqrt{\frac{\Gamma_+}{2}} \ket D \bra G_n \,,\\
    \hat L_\Gamma^{(n)} &= \sqrt{\frac{\Gamma}{2}} \ket G \bra D_n \,,\\
    \hat L_\eta^{(n)} &= \sqrt{\frac{\eta}{2}} \ket{P} \bra{A}_n \,,
\end{align}
describing cavity pumping, cavity decay, pumping of individual pairs, decay of individual pairs back to the state $G$, and relaxation of pairs from $\ket A$ to $\ket F$, respectively.

In a system with dissipation, the spectrum is not characterized by the real eigenenergies of the Hamiltonian Eq.~\eqref{eq:ham}, but it is useful to characterize it by the complex eigenvalues of the non-hermitian Hamiltonian
\begin{align}
    \hat H_{\rm NH} = \hat H - \mi \sum_k \hat L_k^\dagger \hat L_k \,,
    \label{eq:ham_nonherm}
\end{align}
which partly governs the density matrix evolution, as is seen by rewriting Eq.~(\ref{eq:master}) as
\begin{align}
    \partial_t \hat \rho = -\mi \left(\hat H_{\rm NH} \hat \rho - \hat \rho \hat H^\dag_{\rm NH} \right) + 2 \sum_k \hat L_k \hat \rho \hat L_k^\dag. \label{eq:master2}
\end{align}
The real parts of the eigenvalues of $\hat H_{\rm NH}$ correspond to the eigenenergies of the different states in the absence of dissipation, whereas the imaginary parts of the eigenvalues determine the width of the spectral peaks, corresponding to the rate at which one eigenstate decays into other eigenstates. For small dissipation, the eigenstates of the non-hermitian Hamiltonian are close to the eigenstates of Eq.~\eqref{eq:ham}. In contrast, for large dissipation, when the width of the peaks becomes comparable to their separation, different spectral peaks can merge. This happens e.g.~at the transition from strong to weak coupling for increasing $\kappa$, when the difference between the eigenenergies of the two polaritons $\ket +$ and $\ket -$ vanishes, and instead two states with different decay rates, one more photon-like and one more exciton like, are the proper eigenstates of $\hat H_{\rm NH}$. Working with the full master equations and the non-hermitian Hamiltonian allows us to treat both weak and strong coupling in the same formalism.

\subsection{Discussion of possible model implementations}
\label{ssec:nanocrystalnumbers}

We choose our simple toy model in order to study fundamental collective quantum effects by using well established quantum optics tools, such as the adiabatic elimination presented in Sec.~\ref{sec:elim}. Here, we discuss in which limits this simple model can be a realistic model for experiments. In particular, we argue that the main effect observed in this manuscript, i.e.~the cavity-enhanced transfer rates in the presence of dissipation, should also be observable in realistic setups.

The Hamiltonian Eq.~\eqref{eq:ham} has been proposed by Mandal \textit{et al.}~\cite{Mandal_Polari_2020} for an electron transfer setup with a single donor coupled to a cavity. There, it has been proposed that nanocrystal donors and organic molecular acceptors (e.g.~CdS and anthraquinone~\cite{Zhu2014Feb}) could be a specific system to observe cavity modified electron transfer. In this case we identify the following states: $\ket G$ is the total ground state of the system before electron transfer, $\ket D$ corresponds to a state with an excited electron-hole pair in the nanocrystal. $\ket A$ corresponds to the state with an electron in the lowest unoccupied molecular orbital of the organic molecule and a hole remaining on the semiconductor. We consider the state $\ket A$ to have the same nuclear positions as the state $\ket D$, and is thus vibrationally excited. The additional state $\ket F$ corresponds to the vibrational ground state of the charged organic molecule. 

Our setup is a generalization of the setup of Mandal \textit{et al.} to many donor-acceptor pairs, but neglecting internal motion. Instead of a single pair~\cite{Mandal_Polari_2020}, we consider $N=10^4$ pairs, homogeneously coupled to the cavity field, e.g.~by being placed in the central plane of a single-mode \SI{210}{\nano\metre} Fabry-Perot cavity [see Fig.~\ref{fig:setup}(a)]. For such a scenario, precise values of the parameters depend on the specific choices of the type and the size of the donor nanocrystal and the type of the organic molecule acceptor. Typical orders of magnitude are~\cite{Mandal_Polari_2020,Zhu2014Feb,Gupta2013Dec}: $g_c = \SI{0.2}{e\volt}$, $\kappa = \SI{1}{e\volt}$, $\Gamma = \SI{3e-7}{e\volt}$, $\Delta=\SI{0.2}{e\volt}$, $V=\SI{0.1}{e\volt}$, and $\omega_{DG} = \SI{3}{e\volt}$. Note that here we consider the (untypical) case of a strong coherent donor-acceptor coupling strength $V$ outside the regime of validity of Marcus theory $V\ll k_BT$~\cite{Mandal_Polari_2020}. For the internal relaxation in the acceptor, which we artificially summarize in the decay rate $\eta$, we choose a value of \SI{e-2}{e\volt}. We choose it not too large to avoid blocking population transfer into the state $\ket A$ by the continuous Zeno effect~\cite{Misra_TheZe_1977}, and not too small such that population can decay into $\ket F$. Note that the precise choice of $\eta$ does not influence our conclusions on the scaling behavior of transfer rates discussed below. Note further that here we focus on unidirectional electron transfer from donor to acceptor and thus ignore electron transfer back to the donor after the molecule has relaxed into the final state $\ket F$~\cite{Zhu2014Feb}. We also ignore the dynamics of the hole in the semiconductor, which can be important in real experiments~\cite{Zhu2014Feb}, and any temperature induced effects, as the thermal energy at room temperature $k_BT\approx\SI{25}{\milli e\volt}$ is much smaller than any energy scale associated with the Hamiltonian in Eq.~\eqref{eq:ham}. In the following, we discuss possible implementations of the incoherent photon pumping, and the role of motional degrees of freedom.

\subsubsection{Incoherent pumping}
We assume that the experiment takes place in an external, incoherent electromagnetic field, which pumps photons into the system. Such an external field could be e.g.~created by a lamp or sunlight, as long as there is sufficient emission around $\SI{3}{e\volt}$. The light can be either absorbed by the cavity or by the donor-acceptor pairs, which is described by Lindblad operators $\hat L_{\kappa_+}$ and $\hat L_{\Gamma_+}^{(n)}$, respectively. The geometry of the cavity and the incoming light can be used to control the relative magnitude of the two terms: If all incoming light is aimed at the cavity mirrors, only $\hat L_{\kappa_+}$ is non-zero, whereas if the light enters perpendicular to the cavity axis, $\hat L_{\kappa_+}$ will vanish and the $\hat L_{\Gamma_+}^{(n)}$ will be non-zero. In order to compare both processes, in the following we first keep $\Gamma_+ = 10^{-4}\Gamma$ constant and vary $\kappa_+$ [see Fig.~\ref{fig:setup}(e) and \ref{fig:evol}], and later analyze the role of the individual processes (Figs.~\ref{fig:transcontour} and~\ref{fig:transrate-wdark}).

Note that for transitions with an excitation energy $\sim \SI{3}{e\volt}$ as considered here, thermal excitations are negligible because the Boltzmann factor completely vanishes at room temperature $T$, $\exp[-\omega_{DG}/(k_BT)] \sim 10^{-61}$ ($k_B$ is the Boltzmann constant). For experiments that use transitions in the infrared spectrum, or work at higher temperatures, thermal photon populations might also be an incoherent photon source.

\subsubsection{External motion}
For nanocrystal donors, we can consider stationary donor-acceptor pairs in the center of the cavity. However, one may argue that our results may also hold for some more general setups with external motion, such as liquid solutions. One can argue for this since the distances over which the pairs are moving on relevant time-scales are irrelevant, even in a gas phase at room temperature (see argument below). In experiments, collisions with solvent molecules might further suppress these already negligible effects. However, in many experiments solvents play a crucial role in the reaction by drastically modifying the local potential energy landscape. This could e.g.~lead to energetic disorder and a mixing of bright and dark states, which for simplicity we do not consider here.

At room temperature, in a gas phase, donor-acceptor pairs move at velocities $v \sim \sqrt{k_BT/m} \lesssim \SI{160}{\metre\per\second}$, where $m \gtrsim \SI{100}{u}$ is the mass of a donor-acceptor pair, and $T\approx \SI{300}{\kelvin}$ is the temperature. The bright state lifetime is $\sim 1/\kappa \approx \SI{0.66}{\femto\second}$, the dark state lifetime is $\sim 1/\eta \approx \SI{66}{\femto\second}$, during which each pair moves by $\sim v/\kappa \approx \SI{0.1}{\pico\metre}$, or $v/\eta \approx \SI{10}{\pico\metre}$, respectively. This is negligible compared to the wavelength of the cavity mode $\lambda \sim \SI{210}{\nano\metre}$, so that entanglement between motion and electronic degrees of freedom should be suppressed over relevant timescales. Thermal motion also leads to a negligible Doppler broadening of $\sqrt{k_BT/(mc^2)}\omega_{DG} \approx \SI{1.6}{\micro e\volt}$.

\subsubsection{Internal motion}

By internal motion, we mean the relative motion of the nuclei in the donor and the acceptor, i.e.~vibrations, which generally couple to the electronic degrees of freedom. Depending on the type of donors and acceptors, this coupling can be large and lead to important effects such as vibrational decoupling~\cite{herrera2016cavity,zeb2018exact}, conical intersections~\cite{vendrell2018collective,feist2018polaritonic}, or even entanglement between vibrations and electronic degrees of freedom~\cite{delPino2018Nov}. In nanocrystals, however, this coupling is typically small~\cite{Mandal_Polari_2020}. Therefore, for such systems it can be realistic to neglect it. For the organic molecule acceptors we include the effects of internal motion in the effective rate $\eta$. This then allows for a disentangled description, where both internal and external motional degrees of freedom separate from the electronic dynamics. In the following, we will only focus on the electronic dynamics.

\section{Adiabatic Elimination} \label{sec:elim}

Numerically simulating the full master equation Eq.~\eqref{eq:master} becomes impossible for just a few donor-acceptor pairs $N \gtrsim 10$ even for our simplified model, as the number of elements in the density matrix scales as $4^{2N}$. We thus proceed by introducing an adiabatic elimination procedure which allows us to significantly reduce this complexity. Adiabatic elimination is a standard technique in quantum optics~\cite{gardiner2004quantum,Reiter2012Mar,gardiner1985handbook,Finkelstein-Shapiro2020Adiabatic} that can be used if the system can be split into fast and slow evolving subspaces. In this case, the fast subspace is typically close to its equilibrium state. Adiabatic elimination then allows us to eliminate the fast degrees of freedom by approximating them with their equilibrium value, and derive an effective master equation for the slow degrees of freedom only, significantly reducing the complexity of the problem. Here, we choose the photons and the excited states $\ket D$ or $\ket A$ as fast degrees of freedom and the states $\ket G$ and $\ket F$ as slow ones, such that the effective master equation following adiabatic elimination describes direct population transfer from $\ket G$ to $\ket F$ [see Fig~\ref{fig:setup}(b)]. In our case the elimination condition is fulfilled when the excited states decay much faster than they are pumped, and thus $\langle \hat N_e \rangle \ll 1$.

\smallskip

Adiabatic elimination can be formalized by defining projection operators, $\hat P = \bigotimes_n (\ket G \bra G_n + \ket F \bra F_n)$ and $\hat Q = \id - \hat P$, into the slow and fast manifold, respectively, where $\bigotimes_n$ denotes a tensor product over all pairs. The goal is to derive an effective equation of motion for $\hat P \hat \rho \hat P$ under the condition $\langle \hat N_e \rangle \ll 1$. In order to simplify notation, we define superoperators, which are linear operators acting on operators, i.e.~they can be expressed as matrices acting on operators written as vectors. The relevant projection superoperators are $\mathcal P$ and $\mathcal Q = \id_S - \mathcal P$, where $\id_S$ is the superoperator identity. They are defined by $\mathcal P \hat \rho = \hat P \hat \rho \hat P$, and $\mathcal Q \hat \rho = \hat P \hat \rho \hat Q + \hat Q \hat \rho \hat P + \hat Q \hat \rho \hat Q$, respectively. We further define the evolution superoperator $\mathcal L = -\mi [\hat H, \cdot] + \sum_k \mathcal D [\hat L_k]$, such that the master equation Eq.~\eqref{eq:master} can be written as $\partial_t \hat \rho = \mathcal L \hat \rho$. Superoperators can be written as a tensor product of two operators: one acting from the right as a transpose, and one acting from the left that is: $\mathcal O\vec \rho = (\hat O_1 \otimes \hat O_2)\vec \rho \equiv \hat O_2 \hat \rho \hat O_1^T$. In general, a superoperator is the sum of multiple such terms. For example, we can write $\mathcal P = \hat P \otimes \hat P$, $\mathcal Q = \hat P \otimes \hat Q + \hat Q \otimes \hat P + \hat Q \otimes \hat Q$ and
\begin{align}
    \mathcal L &= -\mi \id \otimes \hat H_\mathrm{NH} + \mi \hat H_\mathrm{NH}^* \otimes \id + 2\sum_k \hat{L}_k^* \otimes \hat L_k \,,
\end{align}
where $\hat{O}^* \equiv (\hat O^\dagger)^T$ denotes the element-wise complex conjugate of an operator $\hat O$.

\smallskip

As shown by Finkelstein-Shapiro \textit{et al.}~\cite{Finkelstein-Shapiro2020Adiabatic}, with these definitions we can write the effective master equation as
\begin{align}
    \partial_t (\mathcal P \hat \rho) &= \mathcal L_{\rm eff} (\mathcal P \hat \rho) \\
    \mathcal{L}_{\rm eff} &= \mathcal{PLP} - \mathcal{PLQ}(\mathcal{QLQ})^{-1}\mathcal{QLP}\, , \label{eq:elim}
\end{align}
where the first term in $\mathcal{PLP}$ describes the direct evolution in the slow subspace (i.e.~$\ket G$ and $\ket F$ in our case), the second term $\mathcal{PLQ}(\mathcal{QLQ})^{-1}\mathcal{QLP}$ describes an excitation into the excited subspace, some excited state evolution, followed by a decay back into the slow subspace. The formula Eq.~\eqref{eq:elim} strongly resembles other forms of perturbation theory, and reduces to textbook second order perturbation theory in quantum mechanics~\cite{sakurai1995modern}, if all dissipative terms vanish. In this case $\mathcal{PLQ}$ and $\mathcal{QLP}$ reduce to the perturbation Hamiltonian, and $\mathcal{PLP} + \mathcal{QLQ}$ reduces to the bare Hamiltonian. In the remainder of this section, we evaluate $\mathcal L_{\rm eff}$ by only considering a single excitation for the excited state propagator $(\mathcal{QLQ})^{-1}$. The solution is given below by Eqs.~\eqref{eq:PLP} and \eqref{eq:PLQQLQQLP} and a formula for the instantaneous transfer rate can be cast into the simple analytical form of Eq.~\eqref{eq:transrate}.
    
\medskip

We first discuss the individual contributions to the dynamics governed by Eq.~\eqref{eq:elim}:

The term $\mathcal{PLP}$ describes the dynamics confined to the ground state manifold. In our model this is only loss of population into the excited state due to pumping of the cavity ($\kappa_+$) or via pumping of individual donors ($\Gamma_+$) by the external incoherent light source, leading to
\begin{align}
    \mathcal{PLP} &= -\hat P \otimes \hat P \qty(\frac{\kappa_+}{2} + \frac{\Gamma_+}{2} \hat N_G) \nonumber\\
    &\quad - \qty(\frac{\kappa_+}{2} + \frac{\Gamma_+}{2} \hat N_G) \hat P \otimes \hat P \label{eq:PLP}.
\end{align}\\

The term $\mathcal{QLP}$ describes the population transfer from the ground state to the excited state manifold, due to the pumping terms above:
\begin{align}
    \mathcal{QLP} &= \kappa_+ \qty(\hat a^{\dagger})^* \otimes \hat a^\dagger + \Gamma_+ \sum_n \qty(\ket D \bra G_n)^* \otimes \ket D \bra G_n \label{eq:QLP}.
\end{align}\\

The term $\mathcal{PLQ}$ describes decay from the excited state into the ground state, either due to photon loss from the cavity ($\kappa$), due to spontaneous emission from the molecules back into the ground state ($\Gamma$), or due to relaxation of the acceptor state ($\eta$):
\begin{align}
    \mathcal{PLQ} &= \kappa \hat a^* \otimes \hat a + \Gamma \sum_n \qty(\ket G \bra D_n)^* \otimes \ket G \bra D_n \nonumber \\
    &\quad + \eta \sum_n \qty(\ket{F} \bra{A}_n)^* \otimes \ket{F} \bra{A}_n \label{eq:PLQ}.
\end{align}

The term $\mathcal{QLQ}$ describes the excited state evolution. This includes coherent transfer between donor state $\ket D$ and acceptor state $\ket A$, the coherent energy exchange between donor state $\ket D$ and cavity, as well as population loss due to emission:
\begin{align}
    \mathcal{QLQ} &= -\mi\id \otimes \hat Q \hat H_\mathrm{NH} \hat Q + \mi \hat Q \hat{H}_\mathrm{NH}^* \hat Q \otimes \id \nonumber \\
    &\quad - \mi \hat Q \otimes \hat P \hat H_\mathrm{NH} \hat P + \mi \hat P \hat{H}_\mathrm{NH}^* \hat P \otimes \hat Q \,, \label{eq:QLQ}
\end{align}
with the projections of the non-hermitian Hamiltonian
\begin{align}
    \hat P \hat H_\mathrm{NH} \hat P &=  - \mi \frac{\kappa_+}{2} - \mi \frac{\Gamma_+}{2} \hat N_G \label{eq:PHP}, \\
    \hat Q \hat H_\mathrm{NH} \hat Q &= g \sum_n \qty(\ket D \bra G_n \hat a + \ket G \bra D_n \hat a^\dagger) \nonumber \\
    &\quad + V \sum_n \qty(\ket A \bra D_n + \ket D \bra A_n) \nonumber \\ 
    &\quad + \qty(\Delta - \mi \frac{\eta}{2}) \hat N_A - \mi \frac{\kappa}{2} \hat a^\dagger \hat a - \mi \frac{\Gamma}{2} \hat N_D  \label{eq:QHQ}.
\end{align}
Note that since we only consider a single excitation, there can be neither further pumping from the excited state, nor relaxation into the excited state. \\

We can now directly proceed to diagonalize $\mathcal{QLQ}$ using the eigenstates and complex eigenenergies of $\hat H_{\rm NH}$ discussed above and given in Appendix~\ref{app:eigenstates}. We label the eigenstates $\ket{\mathcal N_F, \psi}$, where $\mathcal N_F = \{i_1, \dots ,i_{N-M}\}$ is the set of pairs in state $\ket F$, $|\mathcal N_F|$ and $M=N-|\mathcal N_F|$ are the numbers of pairs in and outside of state $\ket F$, respectively, and $\psi \in \{G_c, k_\pm, +, -, X\}$ is one of the eigenstates of $\hat H_{\rm NH}$ for given $\mathcal N_F$. The corresponding complex eigenenergies are denoted as $E_\psi^{(M)}$. We label the elements of the excited state propagator $(\mathcal{QLQ})^{-1}$ by $\mathcal G$, with elements
\begin{align}
    & \mathcal G\qty(\mathcal N_F', \psi; \mathcal N_F, \phi) \nonumber \\
    \equiv&\qty(\bra{\mathcal N_F, \psi} \otimes \bra{\mathcal N'_F, \phi}) \qty(\mathcal{QLQ})^{-1} \qty(\ket{\mathcal N_F, \psi} \otimes \ket{\mathcal N'_F, \phi}) \nonumber \\
    =& \Big[\mi \qty(E^{(N-|\mathcal N_F|)}_\psi)^* - \mi E^{(N-|\mathcal N_F'|)}_\phi\Big]^{-1},
    \label{eq:propagator}
\end{align}

Using the expressions $\mathcal{QLP}$ and $\mathcal{PLQ}$ in the diagonal basis of $\hat H_\mathrm{NH}$ given in Appendix~\ref{app:plqqlp}, calculating $\mathcal{PLQ}(\mathcal{QLQ})^{-1}\mathcal{QLP}$ is straightforward. We find:
\begin{widetext}
\begin{align}
    \mathcal{PLQ}(\mathcal{QLQ})^{-1}\mathcal{QLP} =
    \sum_{\substack{\mathcal{N}_F, \mathcal N_F' \\ \psi,\phi \in \{ k_\pm,\pm,X\} }} \Bigg\{ &\qty[\kappa_+ \qty(c_\psi^\mathrm{ph})^*c_\phi^\mathrm{ph} + \Gamma_+ \sum_n \qty(c_\psi^{D_n})^*c_\phi^{D_n}]
    \mathcal G(\mathcal N_F, \psi; \mathcal N_F', \phi) \nonumber\\
    &\quad \times \qty[\kappa \qty(\bar c_\psi^\mathrm{ph})^* \bar c_\phi^\mathrm{ph} + \Gamma \sum_{n'} \qty(\bar c_\psi^{D_{n'}})^* \bar c_\phi^{D_{n'}}]
    \ket{\mathcal N_F, G}\bra{\mathcal N_F, G}\otimes \ket{\mathcal N_F', G}\bra{\mathcal N_F', G} \nonumber \\
    +&\qty[\kappa_+ \qty(c_\psi^\mathrm{ph})^*c_\phi^\mathrm{ph} + \Gamma_+ \sum_n \qty(c_\psi^{D_n})^* c_\phi^{D_n}]
    \mathcal G(\mathcal N_F, \psi; \mathcal N_F', \phi)
    \qty[\eta \sum_{n'} \qty(\bar c_\psi^{A_{n'}})^* \bar c_\phi^{A_{n'}}] \nonumber \\
    &\quad \ket{\mathcal N_F \dot \cup \{n\}, G}\bra{\mathcal N_F, G}\otimes \ket{\mathcal N_F' \dot \cup \{n\}, G}\bra{\mathcal N_F', G} \Bigg\} \label{eq:PLQQLQQLP},
\end{align}
\end{widetext}
where $c^\psi_\phi$ and $\bar c^\psi_\phi$ are state overlaps defined in Appendix~\ref{app:calcc}, and the state notation ``$\mathcal N_F \dot \cup \{n\}$'' indicates that an electron was transferred in pair $n$, bringing it from state $\ket G$ to state $\ket F$.
The first two lines in the right hand side of Eq.~\eqref{eq:PLQQLQQLP} describe processes that do not contribute to population transfer: A photon is injected into the cavity ($\kappa_+$ terms) or a molecule is brought into the donor state ($\Gamma_+$ terms), and then after some coherent evolution the excitation is lost back into the initial state via cavity decay ($\kappa$ terms) or spontaneous emission ($\Gamma$ terms). The third and fourth line describe population transfer of a donor-acceptor pair from the ground state $\ket G$ to the final state $\ket F$: A pair is excited or a photon is absorbed by the cavity, the excitation is transferred to a pair in the acceptor state, and the excitation is lost via the $\hat L_\eta$ channel to reach the final state $\ket F$, as discussed in detail in Sec.~\ref{ssec:rateeq}. Note that no coherences between states with different $\mathcal{N}_F$ are generated by these terms. Thus there will be no coherences between $\ket G$ and $\ket F$ other than those already present in the initial state. For the default case where in the initial state $\mathcal N_F' = \mathcal N_F$, we use the short hand notation for the propagator $\mathcal G(M, \psi, \phi) = \mathcal G\qty(\mathcal N_F, \psi; \mathcal N_F, \phi)$ with $M = N - |\mathcal N_F|$.

\section{Results} \label{sec:results}

\subsection{Rate equation}
\label{ssec:rateeq}

Our effective master equation Eq.~\eqref{eq:elim} with the term of Eq.~\eqref{eq:PLQQLQQLP} describes purely dissipative population transfer from $\ket G$ to $\ket F$. Only the last two lines in Eq.~\eqref{eq:PLQQLQQLP} contribute. The transfer is characterized by the instantaneous transfer rate from states $\ket{G}$ to $\ket{F}$ [see Fig.~\ref{fig:setup}(b)], that is, the situation where $M$ pairs are transiently in state $\ket{G}$, 
\begin{align}
    r = \sum_{\phi, \psi}&\qty[\kappa_+ \qty(c_\psi^\mathrm{ph})^*c_\phi^\mathrm{ph} + \Gamma_+ \sum_n \qty(c_\psi^{D_n})^* c_\phi^{D_n}] \nonumber\\
    &\qquad\times \mathcal G(M, \psi, \phi) \qty[\eta \sum_{n'} \qty(\bar c_\psi^{A_{n'}})^* \bar c_\phi^{A_{n'}}] \label{eq:transrate} \, .
\end{align}
This instantaneous transfer rate $r$, which can be computed efficiently for arbitrary system parameters, is the main result of our paper. 

The rate $r$ in Eq.~\eqref{eq:transrate} is determined by a sum over contributions from the eigenstates of $\hat H_{\rm NH}$, $\psi,\phi \in \{ k_\pm,\pm,X\}$. We identify two independent processes driving population transfer, given by the $\kappa_+$ and $\Gamma_+$ terms, respectively: i) Cavity photons that are incoherently excited at rate $\kappa_+$ are transformed into excitations of $\ket D$ and $\ket A$ via \emph{collective} excitations $\{\pm,X\}$, which is reflected by the state overlaps of those eigenstates with the photon mode, $c^{\rm ph}_\psi$ (see Appendix~\ref{app:calcc} for state overlap definitions). Note that the overlap with the (localized) dark states vanishes $c^{\rm ph}_{k_\pm} = 0$. The second line in Eq.~\eqref{eq:transrate} describes a transfer of the excitation to the final state $\ket{F}$ via the acceptor states $\ket{A}$, which is seen by the overlaps $\bar c^{A_n}_\psi$, and the decay rate $\eta$ for decaying into the final state $\ket F$; 
ii) Alternatively, each molecule in state $\ket G$ can be \emph{individually} excited into the donor states $\ket D$ at rate $\Gamma_+$, which overlaps with all states $\{k_\pm,\pm,X\}$ via $c_\psi^{D_n}$. With the same mechanism as in i) [second line in Eq.~\eqref{eq:transrate}] the excitation can then be transferred into the acceptor state $\ket A$, and finally to the state $\ket F$. 

The ``propagator'' relevant to both processes, is [from Eq.~\eqref{eq:propagator}]:
\begin{align}
    & \mathcal G\qty(M, \psi, \phi) = \Big[\mi \qty(E^{(M)}_\psi)^* - \mi E^{(M)}_\phi\Big]^{-1}
\end{align}
with $E^{(M)}_\psi$ the complex excited state eigenvalues. For $\psi = \phi$, $\mathcal{G}$ reduces to the inverse decay rate, i.e.~the life-time, of the excited state $\psi$. Although all excited states decay fast compared to pumping rates, in the propagator $\mathcal{G}$ transfer via relatively long-lived states is preferred over transfer via states that decay quickly back into the state $\ket G$.

\smallskip

The relative importance of the two processes i) and ii) above depends directly on the relative magnitude of $\kappa_+$ and $\Gamma_+$ [see Fig.~\ref{fig:setup}(d)], on the instantaneous ground state population $M$, and indirectly on the other system parameters via state overlaps $c_\phi^\psi$, and the propagator $\mathcal G$. Note that Eq.~\eqref{eq:transrate} is independent of the total number of pairs $N$, as pairs in the state $\ket F$ do not participate in any dynamics. Fig.~\ref{fig:setup}(e) shows the dependence of the rate $r$ on the collective cavity coupling $g_c = g\sqrt{N}$ for different values of $\kappa_+$. For $\kappa_+ = 0$, only the $\Gamma_+$-channels contribute and the transfer rate is essentially independent of $g_c$. This indicates that the transfer occurs dominantly via dark states, which are unmodified by $g_c$. In contrast, if we increase $\kappa_+$, we find a strong dependence of $r$ on $g_c$. In both extreme cases $g_c=0$ and $g_c \rightarrow \infty$, the $\kappa_+$-channels do not contribute, but the transfer rate $r$ has a maximum for a small $g_c$. The relative importance of the two channels and their dependence on different system parameters are discussed in detail in Subsection~\ref{ssec:discussion}.

\subsection{Time evolution}
\label{ssec:evol}

\begin{figure}
    \centering
    \includegraphics[width=\columnwidth]{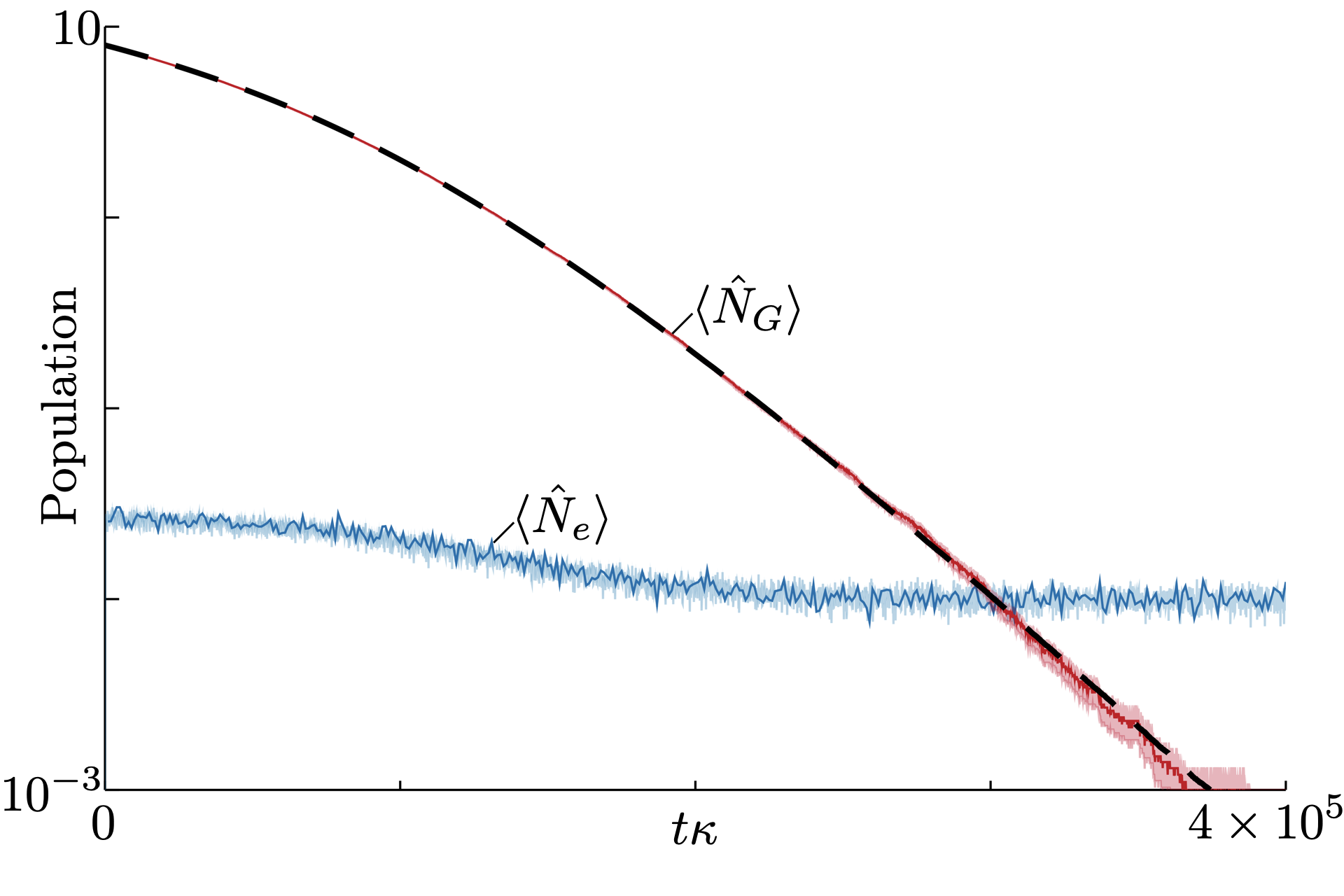}
    \caption{Test of validity for a small system. Comparison of the time evolution (log-log scale) of state populations described by the full master equation (continuous lines, red and blue) with the ones obtained from the effective rate equation Eq.~\eqref{eq:transrate} (dashed black line) for $N=8$. The upper red and dashed black lines show the ground state population $\langle \hat N_G \rangle$, the blue line the excited state population $\langle \hat N_e \rangle$ (which vanishes in the limit of validity of the effective equation). For the full simulations, the shaded area corresponds to two standard deviations of the mean calculated from $10^4$ quantum trajectories. Parameters for the simulations: $g_c = 0.2\kappa$, $\kappa_+ = 10^{-2}\kappa$, $\Gamma = 3\times 10^{-7}\kappa$, $\Gamma_+ = \Gamma/6$, $\Delta=0.2\kappa$, $V=0.1\kappa$, and $\eta=10^{-2}\kappa$.}
    \label{fig:elimver}
\end{figure}

From Eq.~\eqref{eq:transrate}, we can numerically compute the $\ket{G} \to \ket{F}$ transfer time evolution. In particular, we are interested in the evolution of the populations $\langle \hat N_G\rangle$ and $\langle \hat N_F\rangle$ in states $\ket G$ and $\ket F$, respectively. With the rates from Eq.~\eqref{eq:transrate} we can do this easily for very large system sizes $N \sim 10^4$. To do so, we describe the density matrix as a statistical mixture of pure states, each of which can be stochastically evolved in time. Such a technique is well developed and known as quantum trajectories or quantum Monte-Carlo wavefunction method~\cite{daley2014quantum}. Here, such a quantum trajectory algorithm is trivially simplified by the fact that there is no coherence in the effective dynamics, i.e.~$\hat H_{\rm eff} = 0$ (see Appendix~\ref{app:traj}).

\begin{figure}
    \centering
    \includegraphics[width=\columnwidth]{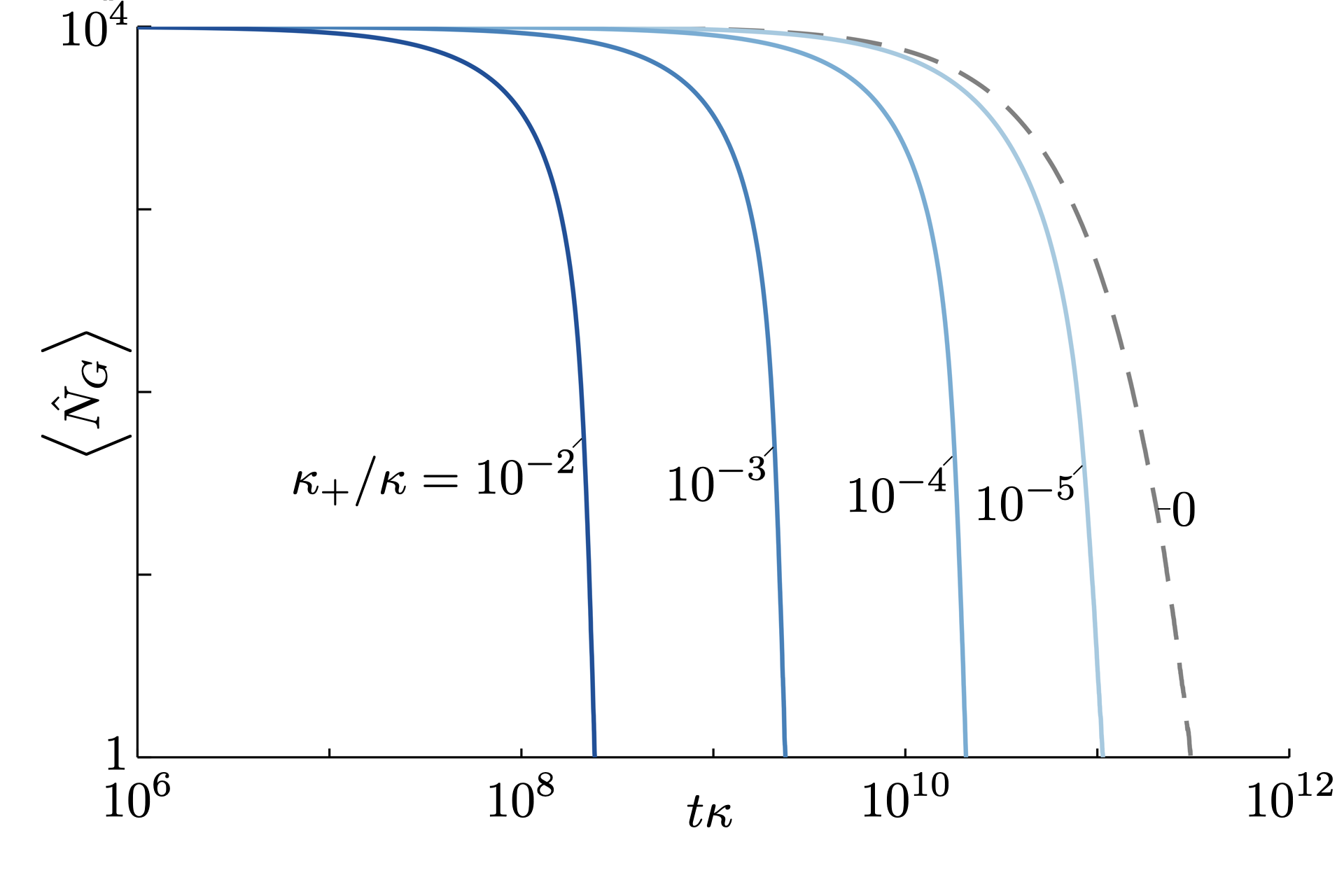}
    \caption{Large system time evolution with initially $N=10^4$ ground-state pairs. Shown is the evolution of $\langle \hat N_G \rangle$ for different cavity pumping rates $0\leq\kappa_+\leq 10^{-2}\kappa$ (log-log scale). We simulate Eq.~\eqref{eq:transrate} for with $g_c = 0.2\kappa$, $\Gamma = 3 \times 10^{-7} \kappa$, $\Gamma_+ = 10^{-4}\Gamma$, $\Delta=0.2\kappa$, $V=0.1\kappa$, and $\eta=10^{-2}\kappa$.}
    \label{fig:evol}
\end{figure}

To check the validity of our adiabatic elimination, we initially compare our results to a numerical simulation of the full master equation~\eqref{eq:master} for a small system. For this simulation we also use a quantum trajectory approach~\cite{daley2014quantum}. In Fig.~\ref{fig:elimver} we compare the results for $N=8$. We find that for typical parameters, the elimination condition $\langle \hat N_e \rangle \ll 1$ is fulfilled and both descriptions match even for the largest considered cavity drive that we consider, $\kappa_+/\kappa = 10^{-2}$. Note that when varying $N$, to make small and large system sizes comparable we approximately match the collective cavity coupling $g_c = \sqrt{N}g$ and pumping rate $N\Gamma_+$ (instead of $g$ and $\Gamma_+$). In this way we achieve an equivalent mixing of photons and excited states and equivalent total incoherent transfer rates.

\begin{figure*}
    \centering
    \includegraphics[width=\textwidth]{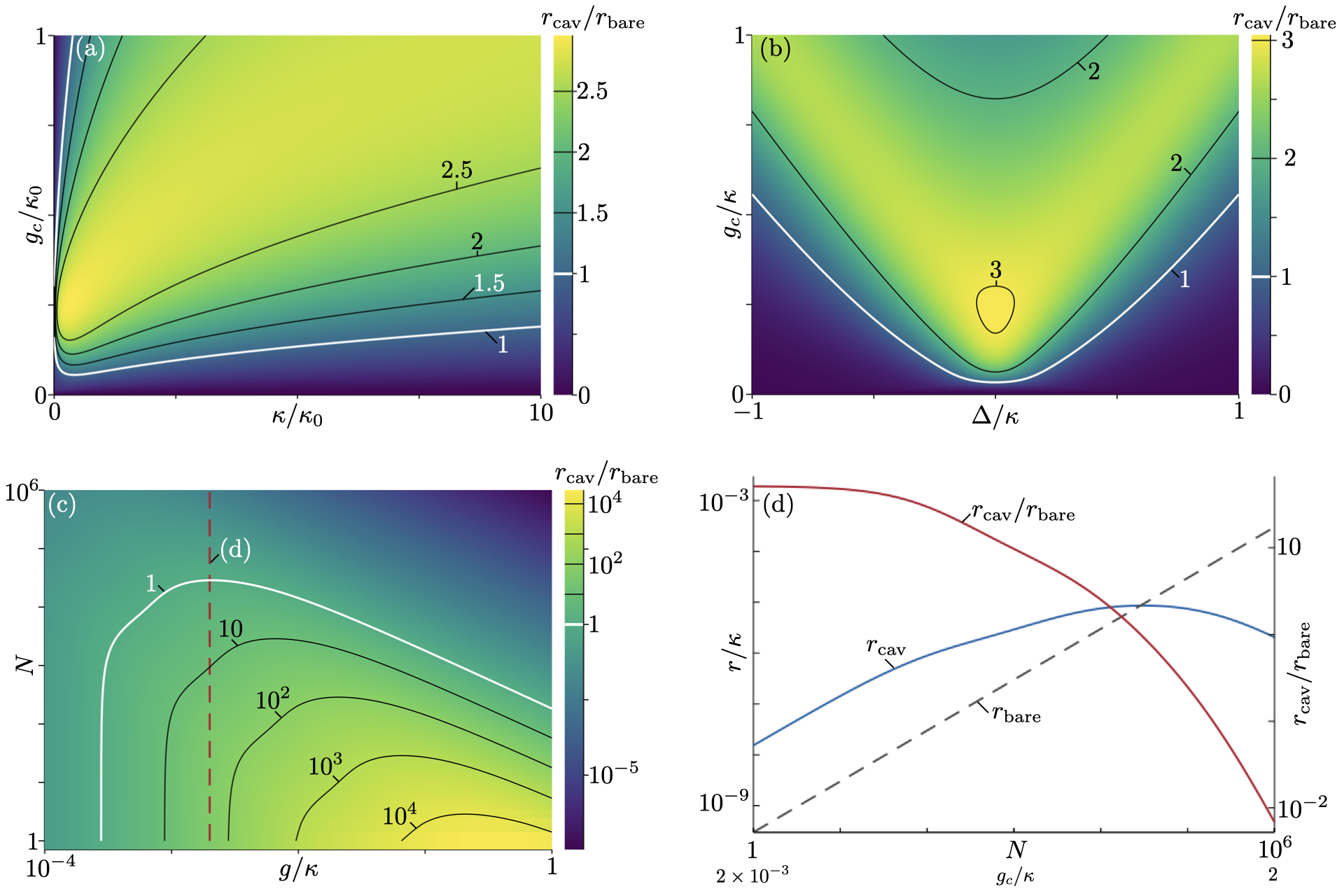}
    \caption{Cavity modification of the transfer rate. (a, b, c) Contour plots of the transfer rate $r_{\rm cav}/r_{\rm bare}$. $r_{\rm cav}$ is given by setting $\kappa_+ = 10^{-3}\kappa$ and $\Gamma_+ = 0$ in Eq.~\eqref{eq:transrate}; $r_{\rm bare}$ is given by setting $\kappa_+ = 0$, $\Gamma_+ = 10^{-3} \Gamma$ and $g=0$ in Eq.~\eqref{eq:transrate} (see text). We define normalized units $\kappa_0$ (e.g.~$\kappa_0=\SI{1}{e\volt}$) and set:  $V=0.1\kappa_0$, $\Gamma=\num{3e-7}\kappa_0$, and $\eta=10^{-2}\kappa_0$. Black and white lines are contours of constant cavity enhancement as indicated in the colorbar, white lines correspond to $r_{\rm cav} = r_{\rm bare}$. (a) $r_\mathrm{cav}/r_\mathrm{bare}$ as a function of $g_c$ and $\kappa$ for $N=10^4$ and $\Delta=0.2\kappa_0$. (b) $r_\mathrm{cav}/r_\mathrm{bare}$ as a function of $g_c$ and $\Delta$ for $N=10^4$ and $\kappa=\kappa_0$.
    (c) $r_\mathrm{cav}/r_\mathrm{bare}$ as a function of $N$ and $g$ for $\Delta=0.2\kappa_0$ and $\kappa=\kappa_0$ (log-scale for all axes and colorbar). (d) $r_\mathrm{cav}$ (blue solid, left axis) and $r_\mathrm{bare}$ (grey dashed, left axis) and $r_\mathrm{cav}/r_\mathrm{bare}$ (red solid, right axis) as a function of $N$, corresponding to a cut (red dashed line) in (c) with $g = 2\times10^{-3}\kappa_0$. The horizontal axis shows $g_c=g\sqrt{N}$ as second scale (log-scale on all axes).}
    \label{fig:transcontour}
\end{figure*}

The algorithm to compute $\langle \hat N_G \rangle$ from Eq.~\eqref{eq:transrate} scales only linearly with the number of pairs, which allows us to consider $N=10^4$ pairs in Fig.~\ref{fig:evol}. In this figure we show the time evolution of the number of ground state donor-acceptor pairs for different values of $\kappa_+$. We observe that a weak cavity pump of only $\kappa_+/\kappa = 10^{-5}$ already leads to a significant acceleration of the ground-state depletion at times when only few ground-state pairs remain coupled to the cavity. While the initial dynamics is barely modified, we find that the long-time depletion becomes even super-exponential. For larger values of $\kappa_+/\kappa$, also the earlier dynamics speeds up. For example increasing $\kappa_+/\kappa$ from $10^{-3}$ to $10^{-2}$, the entire dynamics shifts by a factor of $10$, indicating that the transfer is completely dominated by the cavity in this regime, as the transfer rate $r$ is proportional to the rate at which the cavity is pumped $\kappa_+$.

\subsection{Discussion of the rate equation}
\label{ssec:discussion}

In this subsection, we analyze the dependence of the transfer rate $r$ given by Eq.~\eqref{eq:transrate} on the various model parameters, and identify for which parameters the presence of a cavity increases $r$. Fig.~\ref{fig:transcontour}(a) to (c) are contour plots of the rate modification due to the cavity: $r_{\rm cav} / r_{\rm bare}$, where $r_{\rm cav}$ is given by setting $\Gamma_+=0$ in Eq.~\eqref{eq:transrate}, and $r_{\rm bare}$ is given by setting $\kappa_+=0$ and $g=0$ in Eq.~\eqref{eq:transrate}, describing transfer via the cavity only and without the cavity, respectively. The white lines in Fig.~\ref{fig:transcontour} highlight parameter points where $r_{\rm cav} = r_{\rm bare}$, i.e.~where both setups perform equally well. We set $\kappa_+/\kappa = \Gamma_+/\Gamma = 10^{-3}$ for both scenarios to compare situations with the same driving field: If interaction with the background photon field is the dominant loss and excitation mechanism, we have $\kappa_+/\kappa = \langle \hat n_{\rm ph} \rangle / (1 + \langle \hat n_{\rm ph} \rangle) = \Gamma_+ / \Gamma$, for $\langle \hat n_{\rm ph} \rangle$ the average photon density in the background field. This can be shown by a standard derivation of the master equation~\cite{cohen1998atom}. Here we analyze the experimentally relevant limit (see Sec.~\ref{ssec:nanocrystalnumbers}), where $\kappa \gg \Gamma$, and thus $\kappa_+ \gg \Gamma_+$. Note that in an opposite limit, the bare transfer would dominate over the one via the cavity.

In all contour plots we can see that for vanishing cavity coupling $g_c = \sqrt{N}g \rightarrow 0$, the cavity transfer rate vanishes $r_{\rm cav} \rightarrow 0$. This happens because for $g_c=0$, photons cannot be transformed into donor states $\ket D$, and thus cannot drive any population transfer. Similarly, for diverging coupling $g_c\rightarrow\infty$, $r_{\rm cav}$ also vanishes, because both upper and lower polariton states [i.e.~eigenstates of the second line of Eq.~\eqref{eq:ham}, $\ket{P_\pm}$ in Fig.~\ref{fig:setup}(c)] are far detuned from the state $\ket A$. This suppresses coherent population transfer from the polariton states to $\ket A$. Between both extreme cases, we find an optimal value of $g_c$, for which $r_{\rm cav}$ is maximal and the cavity enhances the transfer. Notably, for a large parameter regime, the optimal $g_c$ lies in the weak coupling regime $g_c \leq \kappa/4$, a regime where the polariton splitting is not observable
\footnote{$g_c \leq \kappa/4$ is the condition for having two eigenstates with different eigenenergies However, the difference can be only spectrally resolved at a larger $g_c$. Note also that we neglected other dissipation $\Gamma, \eta \ll \kappa$, in this approximate condition.}.

For which parameters weak or strong coupling is optimal is e.g.~seen in Fig.~\ref{fig:transcontour}(a), which shows $r_{\rm cav} / r_{\rm bare}$ as a function of $\kappa$ and $g_c$ in normalized units $\kappa_0$. In the limit of small $\kappa$, the enhancement is maximal for a collective coupling strength matching the energy difference between states $\ket{A}$ and $\ket{D}$, $g_c \sim \Delta = 0.2 \kappa_0$. This can be understood from taking the limit $V\to 0$ in the Hamiltonian Eq.~\eqref{eq:ham}. Then, $\ket{A}$ is in resonance with an upper polariton state [$\ket{P_+}$ in Fig.~\ref{fig:setup}(c)] of the Tavis-Cummings part [second line in Eq.~\eqref{eq:ham}]. We thus find that in this strong coupling limit, the enhancement is efficiently mediated by a polaritonic state. In contrast, for larger values of $\kappa$, $r_{\rm cav}$ is maximal for a $g_c$ inside the weak coupling regime. There, although no polariton splitting is present, the mixing between cavity and donor state induced by the non-hermitian Hamiltonian~\eqref{eq:ham_nonherm} can still be important. Thus, there is an excited state with both cavity and $\ket A$ state contribution which can enable coherent transfer, although in many cases the cavity photon decays before this coherent transfer takes place.
For a given $g_c$ there exists an optimal finite $\kappa > 0$. This optimal $\kappa$ is larger than zero, since in the limit $\kappa\rightarrow0$ the cavity cannot be excited by the incoherent pump. In the limit $\kappa\rightarrow\infty$, the cavity photons are lost before they can be absorbed by a donor-acceptor pair. We find that lines of constant enhancement correspond to $g_c \propto \sqrt{\kappa}$, indicating that the proper figure of merit is the collective Purcell factor $g_c^2/\kappa$, which is not tied to weak or strong coupling.

Similar effects can be observed for variable energy difference between the states $\ket D$ and $\ket A$ in Fig.~\ref{fig:transcontour}(b), which shows $r_{\rm cav} / r_{\rm bare}$ as a function of $\Delta$ and $g_c$. For large $\Delta$, the transfer efficiency is maximal for $g_c \approx |\Delta|$. We again attribute these points of efficient populations transfer to respective resonances with the upper and lower polariton states of the Tavis-Cummings part in the Hamiltonian. For larger $g_c \gg |\Delta|$, the large detuning leads to a large suppression of the transfer rate, as both polaritons are out of resonance. In contrast, for small values of $|\Delta|$, $r_{\rm cav}/r_\mathrm{bare}$ is maximal for a small $g_c$ close to the weak coupling regime. Note that we still find significant cavity enhancement inside this weak coupling regime.

In Fig.~\ref{fig:transcontour}(c) we analyze $r_{\rm cav} / r_{\rm bare}$ as function of $g$ and $N$. Here, we find that increasing $N$ decreases the cavity enhancement factor $r_{\rm cav}/r_{\rm bare}$, although it increases the collective cavity coupling $g_c$. Fig.~\ref{fig:transcontour}(d) is a cut through Fig.~\ref{fig:transcontour}(c) for fixed $g$. For increasing $N$, initially the collective effects increase the cavity transfer rate $r_{\rm cav}$. However, this effect competes with an increase of $r_{\rm bare} \propto N$ due to the increasing number of transfer channels. For small $N$, the collective effects compensate for the increasing number of transfer channels, keeping $r_{\rm cav}/r_{\rm bare}$ approximately constant. This can also be seen as vertical sections of contours in Fig.~\ref{fig:transcontour}(c) for small $g$. Further increasing $N$, the collective enhancement becomes weaker and $r_{\rm cav}/ r_{\rm bare}$ decreases. For $g_c \gtrsim (\kappa,\Delta)$, $r_{\rm cav}$ decreases for increasing $N$ due to a too large polariton splitting. Note that here $N$ is the number of pairs in state $\ket G$, such that even if initially the enhancement is only weak, at later times a larger enhancement can be observed [see also Fig.~\ref{fig:evol}].

\begin{figure}
    \centering
    \includegraphics[width=\columnwidth]{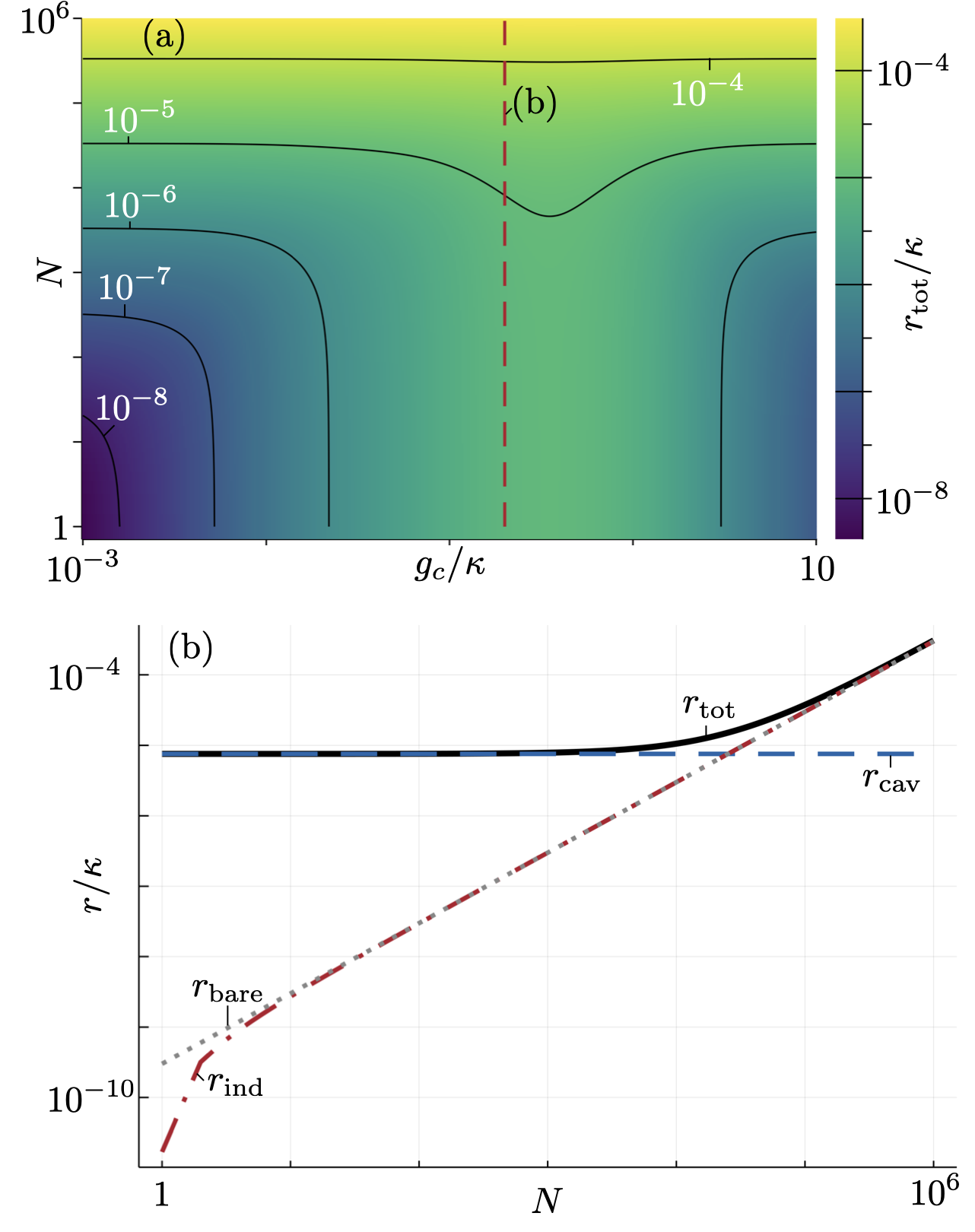}
    \caption{Cavity modification of instantaneous transfer rates including individual pumping [parameters as in Fig.~\ref{fig:transcontour}(c/d), $\Gamma_+ = 10^{-3}\Gamma$]. The total rate $r_\mathrm{tot}$ is the sum of the individual pumping $r_\mathrm{ind}$ [obtained by setting $\kappa_+ = 0$ in Eq.~\eqref{eq:transrate}] and $r_{\rm cav}$. (a) $r_{\rm tot}$, as a function of $g_c$ and $N$ (both axes and color bar are on a log scale). (b) Cut through (a) at the red dashed line $g_c = 0.2\kappa$, and comparison with $r_\mathrm{ind}$ (red dash-dotted) and $r_{\rm ind}$ (blue dashed, log scale axes). $g_c$ is held constant by choosing $g = g_c/\sqrt{N}$.}
    \label{fig:transrate-wdark}
\end{figure}

In the analysis of Fig.~\ref{fig:transcontour} we found that the main features of the cavity modification of initial transfer rates can be understood in terms of resonance conditions of collective bright states, which is included in the state overlaps [compare Fig.~\ref{fig:setup}(c)]. It is important to point out that in general the dynamics can also involve dark states, populated through individual pumping. Therefore, lastly we also analyze the instantaneous rates with both cavity ($\kappa_+ >0 $) and individual pair ($\Gamma_+ > 0$) pumping. To distinguish a scaling in the number of pairs $N$ from simple polaritonic resonance conditions, we now compare rates as function of the collective cavity coupling, $g_c = g \sqrt{N}$ and $N$ in Fig.~\ref{fig:transrate-wdark}. In general, we now find that, depending on $g_c$, for large enough $N$, the overall instantaneous transfer rate $r_{\rm tot}$ exhibits a dependence on $N$~[Fig.~\ref{fig:transrate-wdark}(a)]. This can be explained by a dominance of dark state transfer for large $N$ [as sketched in Fig.~\ref{fig:setup}(d)]. This is exemplified in Fig.~\ref{fig:transrate-wdark}(b), where we compare the contributions to $r_{\rm tot} = r_{\rm cav} + r_\mathrm{ind}$, which besides the cavity rate $r_{\rm cav}$ now also contains the bare individual pumping part $r_\mathrm{ind}$. We define the latter by setting $\kappa_+ =0$ in Eq.~\eqref{eq:transrate}, which is different from $r_{\rm bare}$ which is defined outside the cavity by setting $g=\kappa_+ =0$. Note that in accordance with the discussion above, $r_\mathrm{cav}$ does not depend on $N$ for constant $g_c$, whereas $r_\mathrm{ind}$ grows with increasing $N$. We find that for large enough $N$, $r_{\rm tot}$ is dominated by $r_\mathrm{ind}$. Furtheremore, for large $N$, the cavity influence on the individual pumping vanishes and $r_\mathrm{ind} \approx r_\mathrm{bare}$. This can be understood from the fact that the states excited by $\Gamma_+$ have vanishing polariton contribution [$c^{D_n}_{\pm, X} \rightarrow 0$ as $N \rightarrow \infty$ in Eq.~\eqref{eq:transrate}], and in our model the dark states $\ket{k_\pm}$ decouple from the cavity. Note that in the regime of small $N$, where individual pumping excites a significant fraction of bright states, we find $r_\mathrm{ind} < r_\mathrm{bare}$.

It is worth to re-emphasize that the dynamics is governed by a non-linear rate equation, which implies that the instantaneous rate from Eq.~\eqref{eq:transrate} also depends on time. Then, the time dependence of resonance conditions, together with the dependence on rates on $N$ due to bright and dark state contributions from Fig.~\ref{fig:transrate-wdark}, give rise to non-trivial decay dynamics, as the super-exponential behavior observed in Fig.~\ref{fig:evol}.

Finally, we want to emphasize that we conclude from Fig.~\ref{fig:transcontour}, that for the numbers relevant for nanocrystal setups as described in Sec.~\ref{ssec:nanocrystalnumbers}, the initial transfer rate can be enhanced by a factor $r_{\rm cav} / r_{\rm bare} \sim 2.5$ in the cavity-coupled scenario.

\section{Conclusion \& Outlook} \label{sec:conclusion}

We have presented a Lindblad master equation approach to analyze electron transfer in donor-acceptor pairs coupled to a cavity, in a model where we treat each pair as a 4-level system. Despite neglecting essential mechanisms such as external and internal motional degrees of freedom, such a toy model can help us to discover fundamental mechanics in polaritonic chemistry, using standard tools from quantum optics. By using an adiabatic elimination procedure, we could simplify the problem by eliminating the excited donor and acceptor states, and the cavity, to derive an effective, classical, rate equation for the population transfer between donor and acceptor. This allowed us to gain analytical insight and to perform numerical simulations for very large system sizes. We verified the validity of this approach by benchmarking it for small systems. 

We have found that transfer occurs via two types of channels: One channel is driven by pumping incoherent photons into the cavity and strongly depends on the cavity parameters; the other $N$ channels are driven by individual excitation processes and are essentially unchanged with respect to the no-cavity scenario. Whereas the first channel uses only the three bright states modified by the cavity, the other $N$ channels transfer population via all excited states, and are dominated by the dark states. As the population transfer is purely dissipative and does not require long-lived coherences between different pairs, we expect it to be robust with respect to perturbations such as dephasing, weak static disorder, and weak coupling to external and internal motion. We have simulated the transfer process for large systems and found that the cavity can enhance reaction rates for realistic magnitudes of parameters. This effect is largest for a finite coupling strength between molecules and cavity that falls into the weak coupling regime for strong dissipation. We have analyzed the relevance of the toy-model approximation with regard to recent experimental setups.

It will be interesting to apply the full adiabatic elimination procedure also in the presence of internal and external motional degrees of freedom, and for thermally activated electron transfer. While our effective master equation is purely dissipative and does not build up any coherence, it will be also interesting to investigate the role of entanglement between electronic and motional degrees of freedom, when including the latter. Even without entanglement, it will be interesting to analyze the role of vibrational decoupling on the final transfer rate~\cite{herrera2016cavity,zeb2018exact}.

The effects observed here are similar to those used to enhance electron transfer or electron-hole generation by plasmonic nano-particles~\cite{Atwater2010Mar,Ikeda2011Feb,Zhang2013Mar}. It is an interesting prospect to investigate how the theory developed here may be helpful to understand those types of systems.

As the model presented here uses generic states, our theory can be directly generalized to other types of photo-activated reactions, such as  photoisomerization~\cite{Hutchison_Modify_2012}, long-range energy transfer~\cite{Coles2014Jul,Zhong2016May,Zhong2017Jul} or singlet triplet transitions~\cite{Stranius2018Jun,Takahashi2019Aug,Eizner2019Dec,Yu2020Sep,Polak2020}.

\begin{acknowledgments}
We are grateful to Stefan Sch\"utz, Cyriaque Genet, and Thomas Ebbesen for stimulating discussions. This work is supported by ANR 5 ``ERA-NET QuantERA'' - Projet
``RouTe'' (ANR-18-QUAN-0005-01), LabEx NIE (``Nanostructures in Interaction with their Environment'') under contract ANR-11-LABX-0058 NIE within the Investissement d’Avenir program ANR-10-IDEX-0002-02. D.~W. acknowledges financial support from Agence Nationale de la Recherche (Grant ANR-17-EURE-0024 EUR QMat).  G.~P. acknowledges support from the Institut Universitaire de France (IUF) and the University of Strasbourg Institute of Advanced Studies (USIAS). Research was carried out using computational resources of the Centre de calcul de l'Universit\'e de \mbox{Strasbourg}. 
\end{acknowledgments}

\section*{Data Availability Statement}
The data that support the findings of this study are available from the corresponding author upon reasonable request.

\section*{Copyright}
This article may be downloaded for personal use only. Any other use requires prior permission of the author and AIP Publishing. This article appeared in J. Chem. Phys. \textbf{154}, 054104 (2021) and may be found at \url{https://doi.org/10.1063/5.0037412}.

\appendix

\begin{widetext}

\section{Eigenstates and eigenenergies of $\hat H_{\rm NH}$}
\label{app:eigenstates}

As discussed in the main text, pairs in state $\ket F$ do not participate in the dynamics and can be largely ignored. We label the set of pairs in state $\ket F$ by $\mathcal N_F$. The remaining pairs are labeled $1, \dots, M$.

For a given $\mathcal N_F$, we construct the $2M-2$ dark states of the cavity $\ket{k_\pm}$ with general form
\begin{align}
    \ket{k_\pm} = \sum_{n=1}^M \frac{\exp(-2\pi \mi kn/M)}{\sqrt{M}} \qty(\alpha_{k_\pm} \ket D_n + \beta_{k_\pm} \ket A_n) \, ,
\end{align}
where $k \in \{1, \dots, M-1\}$ is the quasi-momentum, and $\alpha_{k_\pm}$ and $\beta_{k_\pm}$ are the amplitudes of $\ket D$ and $\ket A$, respectively. Due to destructive interference of the different phases of pairs in state $\ket D$, the overall coupling to the cavity vanishes: $(\sum_n \ket{G}\bra{D}_n) \ket{k_\pm} = 0$. The amplitudes $\alpha_{k_\pm}$ and $\beta_{k_\pm}$, as well as the corresponding eigenenergies $E_{k_\pm}^{(M)}$, are given by the components of the normalized eigenvectors and the eigenvalues of the matrix
\begin{align}
    \begin{pmatrix}
    -\mi\frac{\Gamma}{2} & V \\
    V & \Delta - \mi\frac{\eta}{2}
    \end{pmatrix} \, , \label{eqapp:HDA}
\end{align}
respectively, and are independent of $k$. Note that these eigenstates are in general not orthogonal, i.\,e. $\bra{k_+} \ket{k_-} \neq 0$, because the matrix Eq.~\eqref{eqapp:HDA} is not hermitian.

The three remaining eigenstates have the form
\begin{align}
    \ket{\pm/X} = \qty(\alpha_{\pm/X} \ket{G_c}\otimes\ket{1_\mathrm{ph}} + \beta_{\pm/X} \sum_{n=1}^M \frac{\ket{D}_n}{\sqrt{M}} + \gamma_{\pm/X} \sum_{n=1}^M \frac{\ket{A}_n}{\sqrt{M}}) \, .
\end{align}
The prefactors $\alpha_{\pm/X},\beta_{\pm/X},\gamma_{\pm/X}$, and the eigenenergies are given by the components of the eigenvectors and the eigenvalues of the non-hermitian Hamiltonian in the basis $\ket{G_c}\otimes\ket{1_\mathrm{ph}}$, $\ket D_n$, and $\ket A_n$
\begin{align}
    \begin{pmatrix}
    -\mi \frac{\kappa}{2} & \sqrt{M}g & 0 \\
    \sqrt{M}g & -\mi\frac{\Gamma}{2} & V \\
    0 & V & \Delta - \mi\frac{\eta}{2}
    \end{pmatrix} \, , \label{eqapp:Hpmx}
\end{align}
respectively. The eigenvalue with the largest, intermediate, and smallest real part and the respective eigenvectors are assigned to $\ket +$, $\ket X$, and $\ket -$, respectively. The full solution can be easily computed numerically or analytically, but it is rather long and uninsightful, so it is not given here.
\footnote{see \url{https://www.wolframalpha.com/input/?i=\%7B\%7Bd0\%2C+sqrt\%7BM\%7D*g\%2C+0\%7D\%2C+\%7Bsqrt\%7BM\%7D*g\%2C+d1\%2C+V\%7D\%2C+\%7B0\%2C+V\%2C+d2\%7D\%7D} for the full expression}
Note that also these states are not necessarily orthogonal. A complete set of eigenstates of the first excited manifold ($\langle \hat N_e \rangle = 1$) is given by $\ket{\mathcal N_F, \psi}$ with $\psi \in \{k_+, k_-, +, -, X:  k \in \{1, \dots, M-1\}\}$.

\section{State overlaps}
\label{app:calcc}

We define the inverse bras by $\bra{\overline{k_\pm}}$ etc., such that $\bra{\overline{k_\pm}}\ket{k'_\pm} = \delta_{kk'}$ and $\bra{\overline{k_\pm}}\ket{k'_\mp} = 0$. With these definitions, we can rewrite the non-hermitian Hamiltonian as
\begin{align}
    \hat Q \hat H_\mathrm{NH} \hat Q = \sum_{\mathcal N_F, \phi \in \{k_\pm, \pm, X\}} E^{(M)}_\phi \ket{\mathcal N_F, \phi} \bra{\overline{\mathcal N_F, \phi}} \, .
\end{align}

For a fixed $\mathcal N_F$, we define the state overlaps with the notation $\ket{1_{\rm ph}} \equiv \ket{G_c}\otimes \ket{1_\mathrm{ph}}$:
\begin{align}
    c^\mathrm{ph}_\phi &= \bra{\overline{\phi}}\ket{1_\mathrm{ph}}, \label{eq:cph}\\
    c^\mathrm{D_n}_\phi &= \bra{\overline{\phi}}\ket{D}_n ,\\
    c^\mathrm{A_n}_\phi &= \bra{\overline{\phi}}\ket{A}_n ,\\
    \overline{c}^\mathrm{ph}_\phi &= \bra{1_\mathrm{ph}} \ket{\phi}, \\
    \overline{c}^\mathrm{D_n}_\phi &= \bra{D}_n \ket{\phi}, \\
    \overline{c}^\mathrm{A_n}_\phi &= \bra{A}_n \ket{\phi}, \label{eq:cbarAn}
\end{align}
Using that the identity can be written as $\id = \sum_\phi \ket{\phi} \bra{\overline \phi}$, we can use the $c^\psi_\phi$ to expand the states $\ket{1_\mathrm{ph}}$, $\ket D_n$, and $\ket A_n$ in the eigenbasis of $\hat H_{\rm NH}$
\begin{align}
    \ket{1_\mathrm{ph}} &= \sum_\phi c_\phi^\mathrm{ph} \ket{\phi} \, ,\\ \label{eq:1ph}
    \ket{D}_n &= \sum_\phi c_\phi^{D_n} \ket{\phi} \, , \\
    \ket{A}_n &= \sum_\phi c_\phi^{A_n} \ket{\phi} \, , \\
    \bra{1_\mathrm{ph}} &= \sum_\phi \overline{c}_\phi^\mathrm{ph} \bra{\overline{\phi}} \, , \\
    \bra{D}_n &= \sum_\phi \overline{c}_\phi^{D_n} \bra{\overline{\phi}} \, , \\
    \bra{A}_n &= \sum_\phi \overline{c}_\phi^{A_n} \bra{\overline{\phi}}\, . \label{eq:AnTr}
\end{align}
The sums run over all eigenstates of the first excited manifold of $\hat H_{\rm NH}$. These identities are useful when computing $\mathcal{PLQ}$ and $\mathcal{QLP}$.

The indices $c^\mathrm{ph}_{+/X/-}$ can be computed as the top row of
\begin{align}
    \begin{pmatrix}
    \alpha_+ & \beta_+ & \gamma_+ \\
    \alpha_X & \beta_X & \gamma_X \\
    \alpha_- & \beta_- & \gamma_-
    \end{pmatrix}^{-1} \label{eq:trafopmx} \, .
\end{align}
In order to calculate the $c^{D_n,A_n}_\psi$ terms, we first calculate its Fourier transform $\tilde c^{D_k,A_k}_\psi = \sum_n \exp(-2\pi\mi kn/M) c_\psi^{D_n,A_n} / M$, where $\psi \in \{k_+, k_-, +, -, X\}$. The terms $\tilde c^{D_0,A_0}_+$, $\tilde c^{D_0,A_0}_X$, and $\tilde c^{D_0,A_0}_-$ are obtained by dividing the second, third row of the result of Eq.~\eqref{eq:trafopmx} by $\sqrt{M}$. The $k\neq0$ indices can be computed as
\begin{align}
    \begin{pmatrix}
    \tilde c^{D_k}_{k_+} & \tilde c^{D_k}_{k_-} \\
    \tilde c^{A_k}_{k_+} & \tilde c^{A_k}_{k_-}
    \end{pmatrix} 
    =
    \begin{pmatrix}
    \frac{\alpha_{k_+}}{\sqrt{M}} & \frac{\beta_{k_+}}{\sqrt{M}} \\
    \frac{\alpha_{k_-}}{\sqrt{M}} & \frac{\beta_{k_-}}{\sqrt{M}}
    \end{pmatrix} ^{-1} \,. \label{eq:trafokpm}
\end{align}
Note that due to quasi-momentum conservation, all cross-terms $\tilde c^{D_k,A_k}_{k'_-}$ with $k \neq k'$ vanish. Finally, we can compute $c^{D/A_n}_\psi$ by Fourier transforming
\begin{align}
    c^{D_n,A_n}_\psi = \sum_{k=0}^M \exp(2\pi\mi kn/M) \tilde c^{D_k,A_k}_\psi
\end{align}

The $\bar c^\mathrm{ph}_{+,-,X}$ indices are directly given by the coefficients $\alpha_{+,-,X}$. To compute $\bar c^{D_n,A_n}_\psi$, we again define its Fourier transform $\tilde{\bar c}^{D_k,A_k}$, which is given by $\beta_\psi/\sqrt{M}$ and $\gamma_\psi/\sqrt{M}$ for $\psi \in \{+,X,-\}$, or $\alpha_\psi/\sqrt{M}$ and $\beta_\psi/\sqrt{M}$ for $\psi \in \{k_+, k_-\}$.

The computation of the transfer rate Eq.~\eqref{eq:transrate} is simplified by the straightforward to proof identities
\begin{align}
    &\sum_{\substack{\psi\in\{+,-,X\} \\ \phi\in\{k_+,k_-:k\in\{1,\dots,M-1\}\}}} \qty[\kappa_+ \qty(c_\psi^\mathrm{ph})^*c_\phi^\mathrm{ph} + \Gamma_+ \sum_n \qty(c_\psi^{D_n})^* c_\phi^{D_n}] \mathcal G(M, \psi, \phi) \qty[\eta \sum_{n'} \qty(\bar c_\psi^{A_{n'}})^* \bar c_\phi^{A_{n'}}] \nonumber \\
    +&\sum_{\substack{\psi\in\{k_+,k_-:k\in\{1,\dots,M-1\}\} \\ \phi\in\{+,-,X\}}} \qty[\kappa_+ \qty(c_\psi^\mathrm{ph})^*c_\phi^\mathrm{ph} + \Gamma_+ \sum_n \qty(c_\psi^{D_n})^* c_\phi^{D_n}] \mathcal G(M, \psi, \phi) \qty[\eta \sum_{n'} \qty(\bar c_\psi^{A_{n'}})^* \bar c_\phi^{A_{n'}}]
    = 0 \, ,
\end{align}
\begin{align}
    &\sum_{\substack{\psi\in\{k_+,k_-:k\in\{1,\dots,M-1\}\} \\ \phi\in\{k_+,k_-:k\in\{1,\dots,M-1\}\}}} \qty[\kappa_+ \qty(c_\psi^\mathrm{ph})^*c_\phi^\mathrm{ph} + \Gamma_+ \sum_n \qty(c_\psi^{D_n})^* c_\phi^{D_n}] \mathcal G(M, \psi, \phi) \qty[\eta \sum_{n'} \qty(\bar c_\psi^{A_{n'}})^* \bar c_\phi^{A_{n'}}] \nonumber \\
    =&\sum_{i,j\in\{+,-\}} M^2 (M-1) \Gamma_+ \qty(\tilde c_{k_i}^{D_k})^* \tilde c_{k_j}^{D_k} \mathcal G(M, k_i, k_j) \eta \qty(\tilde{\bar{c}}_{k_i}^{A_{k}})^* \tilde{\bar{c}}_{k_j}^{A_{k}} \, .
\end{align}

\section{$\mathcal{PLQ}$ and $\mathcal{QLP}$ in the eigenbasis of $\hat H_{\rm NH}$}
\label{app:plqqlp}

In order to rewrite $\mathcal{QLP}$ and $\mathcal{PLQ}$ in the eigenbasis of $\hat H_\mathrm{NH}$, we use definitions Eqs.~\eqref{eq:1ph} to \eqref{eq:AnTr}. Dropping the index $\mathcal N_F$ for convenience, we find
\begin{align}
    \mathcal{QLP} &= \kappa_+ \left\{ \qty[\qty(c_+^\mathrm{ph})^*\ket{+} + \qty(c_-^\mathrm{ph})^*\ket{-} + \qty(c_X^\mathrm{ph})^*\ket{X}] \bra{G_c} \otimes \qty[c_+^\mathrm{ph}\ket{+} + c_-^\mathrm{ph}\ket{-} + c_X^\mathrm{ph}\ket{X}] \bra{G_c}\right\} \nonumber \\
    &\quad + \Gamma_+ \sum_n \left[\sum_{\psi \in \{k_\pm, +, -, X\}} \qty(c^{D_n}_\psi)^* \ket \psi \bra{G_c} \otimes \sum_{\phi \in \{k_\pm, +, -, X\}} c^{D_n}_\phi \ket \phi \bra{G_c} \right] \label{eq:QLPeigbas} \, ,
\end{align}

\begin{align}
    \mathcal{PLQ} &= \kappa \left\{ \ket{G_c} \qty[\qty(\bar c_+^\mathrm{ph})^*\bra{+} + \qty(\bar c_-^\mathrm{ph})^*\bra{-} + \qty(\bar c_X^\mathrm{ph})^*\bra{X}] \otimes \ket{G_c}\qty(\bar c_+^\mathrm{ph}\bra{+} + \bar c_-^\mathrm{ph}\bra{-} + \bar c_X^\mathrm{ph}\bra{X}) \right\} \nonumber \\
    &\quad + \Gamma \sum_n \left[\sum_{\psi \in \{k_\pm, +, -, X\}} \qty(\bar c^{D_n}_\psi)^* \ket{G_c}\bra \psi \otimes \sum_{\phi \in \{k_\pm, +, -, X\}} \bar c^{D_n}_\phi \ket{G_c} \bra \phi \right] \nonumber \\
    &\quad + \eta \sum_n \left[\sum_{\psi \in \{k_\pm, +, -, X\}} \qty(\bar c^{A_n}_\psi)^* \ket{F_n} \bra \psi \otimes \sum_{\phi \in \{k_\pm, +, -, X\}} \bar c^{A_n}_\phi \ket{F_n} \bra \phi \right] \label{eq:PLQeigbas} \, ,
\end{align}
where we used the notation $\ket{F_n} \equiv \ket{\mathcal N_F \dot \cup \{n\}, G_c}$.

\section{Quantum trajectories for full dynamics}
\label{app:traj}

Quantum trajectories offer an efficient way to simulate quantum open system dynamics~\cite{daley2014quantum}, by reducing the complexity of simulating the time evolution of the full density matrix to simulating the time evolution of an ensemble of states, and reconstructing the density matrix from this ensemble. In the quantum trajectories algorithm, each state is evolved by the non-hermitian Hamiltonian $\hat H_{\rm NH} = \hat H - \mi \sum_k \hat L_k^\dagger \hat L_k$, according the the modified Schrödinger equation
\begin{align}
    \partial_t \ket \psi = - \mi \hat H_{\rm NH} \ket \psi \, . \label{eq:nhschroed}
\end{align}
The state $\ket \psi$ looses norm in this evolution, and when its norm $\bra \psi \ket{\psi}$ drops below $p$, a uniformly distributed random variable between $0$ and $1$, a quantum jump occurs. A random Lindblad operator $\hat L$ is chosen for the jump, where each Lindblad operator $\hat L_i$ has probability 
\begin{align}
    p_i = \frac{\bra \psi \hat L_i^\dagger \hat L_i \ket \psi}{\sum_k \bra \psi \hat L_k^\dagger \hat L_k \ket \psi} \,.
\end{align}
The jump is then executed by computing
\begin{align}
    \ket{\psi'} = \frac{\hat L_i \ket \psi}{\sqrt{\bra \psi \hat L_i^\dagger \hat L_i \ket \psi}} \, .
\end{align}
Identifying this with the master equation dynamics in Eq.~\eqref{eq:master2}, Eq.~\eqref{eq:nhschroed} corresponds to the first part, and the jumps correspond to the second part.

\bigskip

As the state $\ket F$ does not take part in any dynamics, we describe each pair as a 3-level system with levels $\ket G$, $\ket D$, and $\ket A$. We treat emission to state $\ket F$ as effective pair loss, and compute a new Hamiltonian and Lindblad operators each time a pair is ``lost''. We consider a maximum of 1 cavity photon, which we find to be sufficient. If the system is in its ground state $\ket{G_c}$, there is no coherent evolution $\hat H \ket{G_c} = 0$, and we can compute the decay rate $l = \bra{G_c} \sum_k \hat L_k^\dagger \hat L_k \ket{G_c}$, and calculate the time until the next jump analytically $T = - \ln(p)/l$. To compute the excited state evolution, we compute 
\begin{align}
    \ket{\psi (t+dt)} = \exp(-\mi \hat H_{\rm NH} dt) \ket{\psi (t)}
\end{align}
for a small timestep $dt = 0.01/\kappa$, and calculate the norm of $\bra{\psi} \ket{\psi}$ after every timestep.

\section{Quantum trajectories for effective dynamics}

In order to simulate a trajectory defined by the instantaneous decay rate $r(M)$, which depends only on the number of pairs in state $\ket G$, we can use the method described above for evolution out of the ground state. For a random number $p$, we can compute the time until the first jump as $t_1 = -\ln(p)/r(N)$. For each jump exactly one pair is transferred from state $\ket G$ into state $\ket F$. A full trajectory can thus be calculated from a set of random numbers $p_1, \dots, p_N$ according to $t_i = t_{i-1} - \ln(p_i)/r(N-i+1)$. Between timestep $t_i$ and $t_{i+1}$, exactly $i$ pairs are in state $\ket F$, and $N-i$ pairs are in state $\ket G$. The state is efficiently stored as the number $M$ of ground state pairs. The average and standard deviation are computed by sampling over many trajectories.

\end{widetext}

\section*{References} 
\bibliography{main}

\end{document}